\documentclass[lettersize,journal]{IEEEtran}
\usepackage{mathtools}
\usepackage{newtxmath}
\usepackage{cite}
\usepackage{algorithm}

\newcommand{\mertstitle}{OFDM-RSMA: Robust Transmission under Inter-Carrier Interference}

\newcommand{\diag}[1]{\operatorname{diag}\left({#1}\right)}

\makeatletter

\makeatletter
\def\IEEElabelanchoreqn#1{\bgroup
\def\@currentlabel{\p@equation\theequation}\relax
\def\@currentHref{\@IEEEtheHrefequation}\label{#1}\relax
\Hy@raisedlink{\hyper@anchorstart{\@currentHref}}\relax
\Hy@raisedlink{\hyper@anchorend}\egroup}
\makeatother

\usepackage{dblfloatfix}
\usepackage{balance}
\usepackage[caption=false,font=footnotesize]{subfig}

\usepackage{amsmath,amsfonts}
\usepackage{algorithmic}
\usepackage{graphicx}
\usepackage{textcomp}
\usepackage{xcolor}
\usepackage[short,c2,nocomma]{optidef}

\usepackage{acronym}
\input{acros}
\usepackage{textcomp}

\usepackage{siunitx}
\DeclareSIUnit{\belmilliwatt}{Bm}
\DeclareSIUnit{\dBm}{\deci\belmilliwatt}
\DeclareSIQualifier{\isotropic}{i}
\DeclareSIQualifier{\carrier}{c}

\makeatother

\usepackage{listings}
\def\BibTeX{{\rm B\kern-.05em{\sc i\kern-.025em b}\kern-.08em
		T\kern-.1667em\lower.7ex\hbox{E}\kern-.125emX}}
\begin{document}	
	\title{\mertstitle}

	\author{Mehmet Mert \c{S}ahin\IEEEmembership{, Graduate Student Member IEEE}, Onur Dizdar\IEEEmembership{, Member, IEEE}, Bruno Clerckx\IEEEmembership{, Fellow, IEEE}, Huseyin Arslan\IEEEmembership{, Fellow, IEEE}
    \vspace{-0.8cm}
    \thanks{
    M.M. \c{S}ahin was with the Department of Electrical Engineering, University of South Florida, Tampa, FL, 33620. He is now with Standards and Mobility Innovation Laboratory, Samsung Research America, Plano, TX, 75023, USA (e-mail: m.sahin@samsung.com). 

    O. Dizdar was with the Imperial College of London, SW7 2AZ London, U.K. He is now with VIAVI Solutions UK Ltd., SG1 2AN Stevenage, U.K. (e-mail: onur.dizdar@viavisolutions.com).
    
    B. Clerckx is with the Department of Electrical and Electronic Engineering, Imperial College London, London, SW7 2AZ, U.K. and with Silicon Austria Labs (SAL), Graz A-8010, Austria (e-mail: b.clerckx@imperial.ac.uk; bruno.clerckx@silicon-austria.com)

    H. Arslan is with the Department of Electrical and Electronics Engineering, Istanbul Medipol University, Beykoz, 34810 Istanbul, Turkey (e-mail: huseyinarslan@medipol.edu.tr).
    }
    
    }
    
\maketitle
\thispagestyle{plain}
\pagestyle{plain}

\begin{abstract}
\Ac{RSMA} is a multiple access scheme to mitigate the effects of the \ac{MUI} in multi-antenna systems. 
In this study, we leverage the interference management capabilities of \ac{RSMA} to tackle the issue of \ac{ICI} in \ac{OFDM} waveform. We formulate a sum-rate maximization problem to find the optimal subcarrier and power allocation for downlink transmission in a two-user system using \ac{RSMA} and \ac{OFDM}. A weighted minimum mean-square error (WMMSE)-based algorithm is proposed to obtain a solution for the formulated non-convex problem. We show that the marriage of \ac{RS} with \ac{OFDM} provides complementary strengths to cope with peculiar characteristic of wireless medium and its performance-limiting challenges including \ac{ISI}, \ac{MUI}, \ac{ICI}, and \ac{INI}.  The sum-rate performance of the proposed \ac{OFDM}-\ac{RSMA} scheme is numerically compared with that of conventional \ac{OFDMA} and \ac{OFDM}-\ac{NOMA}. It is shown that the proposed \ac{OFDM}-\ac{RSMA} outperforms \ac{OFDM}-\ac{NOMA} and \ac{OFDMA} in diverse propagation channel conditions owing to its flexible structure and robust interference management capabilities. 
    \acresetall
\end{abstract}	


\vspace{-0.5cm}
\section{Introduction} \label{sec:Introduction}

Over the last two decades, the traffic in \ac{MBB} networks has increased 50-70\% each year all around the world \cite{rohdeSchwarz_THz6G}. Immense traffic growth results in the need for the capacity growth to maintain the quality of service. 
Compared to state-of-the-art wireless networks, next-generation wireless networks are expected to achieve significantly higher capacity, extremely low latency, ultra-high reliability, as well as massive and ubiquitous connectivity in order to meet diverse innovative applications, such as \ac{VR}, \ac{AR}, holographic communication, digital replica and \ac{NTN} networks  \cite{samsung2021_6GWhitePaper}. Moreover, the evolution toward next-generation wireless networks requires a paradigm shift from the communication-oriented design to a multi-functional design, including communication, sensing, imaging, computing, and highly accurate positioning capabilities with mobility. To meet these requirements of diverse applications, advanced multiple accessing schemes capable of supporting massive numbers of users are needed \cite{rsmaSurvey_2022}.  
\vspace{-0.2cm}
\subsection{Background and Related Works}

It has been shown that orthogonal transmission schemes are vulnerable to interference which destroy the orthogonality of the signalling scheme, such as, \ac{ISI}, \ac{ICI}, \ac{MUI}, \ac{ACI}, and \ac{INI}  \cite{bottomley_2011, kihero2019INI, arslan2021OFDMchapter}. Moreover, orthogonal transmission schemes can support a number of users limited by the given orthogonal resources \cite{rebhi2021_SCMAsurvey}. Consequently, non-orthogonal transmission strategies have attracted interest from academia and industry for many years to meet the demand on explosively growing number of devices in \ac{NGMA}. 

Since 2G mobile systems, different non-orthogonal transmission schemes are studied to circumvent the limitations encountered in the orthogonal transmission \cite{Wang2018NOMAstandard,Chen2018NOMAstandard}. Non-orthogonal spreading sequences are utilized in the form of IS-95, \ac{WCDMA}, and \ac{CDMA}2000, and serve as the core technologies of 2G and 3G. Moreover, \ac{MUST} has been considered as a candidate multiple access scheme in 4G LTE-A, where messages of cell-edge and cell-interior users are proposed to be superimposed on the same resource element \cite{Lee2016MUST}. \ac{MUST} scenario guarantees the decodability of messages exploiting substantial power difference between users. Another non-orthogonal transmission strategy that has gained popularity in recent years is called \ac{NOMA}, which is studied in two forms, \ac{PD-NOMA} and \ac{CD-NOMA} \cite{Budhiraja2021_NOMAsurvey}. \ac{PD-NOMA} serves multiple users in the same time-frequency resource block, and separates them in the power domain using \ac{SC} and \ac{SIC} techniques \cite{NOMAbook}. \ac{PD-NOMA} allows superposition of users' messages for transmission, and the \ac{MUI} ensuing from this non-orthogonal transmission is decoded and removed via \ac{SIC} at the receiver. In single-antenna systems, e.g. \ac{SISO} \acp{BC}, \ac{PD-NOMA} has been shown to achieve higher \ac{SE} than \ac{OMA} and simultaneously serve larger number of users at an additional cost of increased transceiver complexity \cite{Dai2018NomaSurvey,islam2017PDNOMAsurvey}. 
\ac{PD-NOMA} has been considered as a study-item in 3GPP in the context of 5G standardization process, however, it is not included in the final 5G releases due to their drawbacks in terms of receiver complexity, weakness in mobility, and its performance being dependent on channel strength disparity between users \cite{behrooz2020NOMAproblems}. Furthermore, the design principle of forcing a user to fully decode the interference from other users limits the advantages of \ac{PD-NOMA} \cite{clerckx_2022PartI}. To overcome the drawbacks of \ac{NOMA}, several schemes are proposed, such as, waveform-domain \ac{NOMA}, which is based on the coexistence of different waveforms dedicated to different applications \cite{sahin2020WaveformDomainNOMA, sahin2021AppBasedNOMA, tusha2020NOMAwOFDMIM}. It is shown that waveform-domain \ac{NOMA} provides significant performance improvement in the power-balanced scenario where conventional \ac{PD-NOMA} is subject to ambiguity region, resulting in a high performance loss.

Stemming from Han-Koboyashi scheme \cite{hanKobayashi_RS_1981}, a recently introduced multiple access scheme named \ac{RSMA} has been shown to encapsulate and surpass the performance of \ac{SDMA}, \ac{NOMA}, \ac{OMA}, and physical layer multicasting in multiple antenna networks in terms of spectral and energy efficiency, latency, and resilience to mixed-critical quality of service \cite{mao_clerckx_li_2018,rsmaSurvey_2022}. \ac{RSMA} is a flexible multiple access scheme which provides robust management of interference by partially decoding the interference and partially treating the remaining interference as noise. \ac{RSMA} splits the user messages into common and private parts, and encodes the common parts into one stream while encoding the private parts into separate streams. The obtained streams are precoded, superposed and transmitted over the wireless channel. Each user first decodes the common stream, then performs \ac{SIC}, and decode their respective private streams. Each receiver reconstructs its original message from the part of its message embedded in the common stream and its intended private stream. 

The conceptual demonstration of \ac{RSMA} in \ac{MU} \ac{MISO} can be seen in Fig. \ref{fig:beamRSMAscheme}, where the common stream is applied beamforming to cover both users and $k$-th private stream is directed towards $k$-th user. To maximize the  achievable sum-rate, the power allocation and precoder design for the common and private streams need to be optimized  \cite{joudeh_2016_SRmax_RSMA, clerckx_2020_MAcomparison, matthiesen2022globalOpRSMA, mishra2022rsmaDL}. 
\ac{RSMA} has been shown to provide unique benefits, such as, enhanced spectral and energy efficiency, multiplexing gain, universality by generalizing conventional \ac{OMA}, \ac{SDMA}, \ac{NOMA}, physical-layer multicasting; flexibility by coping with any interference levels, network loads, services, traffic, user deployments; robustness to inaccurate \ac{CSI} at both transmitter and receiver ends and resilience to mixed-critical quality of service; reliability under short channel codes and low latency \cite{clerckx_2023_rsmaMyths,clerckx2021OJCSmaTech,Mao2019RSspecEEeff,xu2022finiteBlockLengthRSMA,reifert2022comebackKid}.  

\Ac{OFDM} waveform has been widely studied and deployed in wireless communication standards such as 4G-LTE, 5G-NR and Wi-Fi owing to its low-complexity implementation and robustness against frequency selective channels \cite{ankarali2017waveform}. However, `sinc' shaped subcarriers of OFDM makes it vulnerable to \ac{ICI} from various causes, such as, phase noise, Doppler spread, mismatch in local oscillators of receiver and transmitter. Since \ac{ICI} destroys the orthogonality of \ac{OFDM} subcarriers, it results in saturation in data rate and error floor in BER with increasing system power \cite{tiejun2006OFDMuDopp}. 

A method to combat the affects of \ac{ICI} is to use flexible \ac{OFDM} symbol duration and \ac{SCS}. Hence, mixed numerologies are employed in 5G to cater for different deployment scenarios and core services \cite{5G_Dahlman_2021}. Although this approach is efficient to provide the required flexibility, another capacity-limiting factor, \ac{INI}, is introduced into the system. Mixed numerologies cause both the loss of orthogonality among
subcarriers of different numerologies and the difficulty in achieving symbol
alignment in time domain \cite{kihero2019INI}.

\vspace{-0.3cm}
\subsection{Motivation and Our Contributions}
Several works can be found on the analysis of \ac{RSMA} in multicarrier systems \cite{DeCastro2022_RSMAmultiCarrier,Li2020_rsmaMultiCarrier,Chen2020_rsmaMultiCarrier,Chen2021_rsmaMultiCarrier,dizdar2022ComJamRS}. A resource allocation algorithm to maximize the sum-rate for RSMA-based multicarrier system is proposed in \cite{DeCastro2022_RSMAmultiCarrier} by performing user matching, subcarrier assignment and power allocation among subcarriers. Similarly, a three step resource allocation scheme is proposed in \cite{Li2020_rsmaMultiCarrier}, where power allocation on a single subcarrier, matching between user and subcarrier and power allocation among different subcarriers are solved in steps to maximize the sum-rate. \Ac{RS}-based precoding in the overloaded multicarrier multi-group multicast downlink scenario is studied in \cite{Chen2020_rsmaMultiCarrier}, where a joint max-min fairness and sum-rate optimization problems are formulated. The practical implementation of the proposed algorithm is verified via link-level simulations in \cite{Chen2021_rsmaMultiCarrier} by evaluating the \ac{BER} performance. An application of \ac{RSMA} for joint communications and jamming with a multicarrier waveform in \ac{MISO}-\ac{BC} is studied in \cite{dizdar2022ComJamRS}.  \ac{RS}-based precoders are designed to perform simultaneous communications with secondary users and jamming of adversarial users. In abovementioned works, \ac{RSMA} has been considered to address the problems of various systems employing multicarrier waveforms, the problems of the multicarrier waveform itself and its interaction with the wireless propagation channel, {\it i.e.,} the determination of \ac{SCS} to manage \ac{ICI} under mobility, are not addressed. 

\begin{figure}[t]
    \vspace{-0.2cm}
    \centering
    \subfloat[]{
    \includegraphics[width=0.6\linewidth]{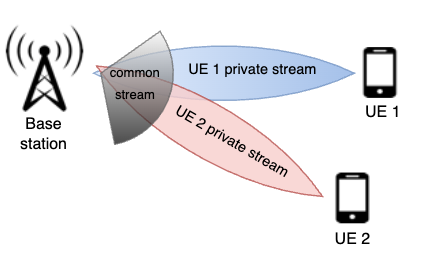}
    \label{fig:beamRSMAscheme}} \vspace{-0.1cm} \hfill
    \subfloat[]{
    \includegraphics[width=0.9\linewidth]{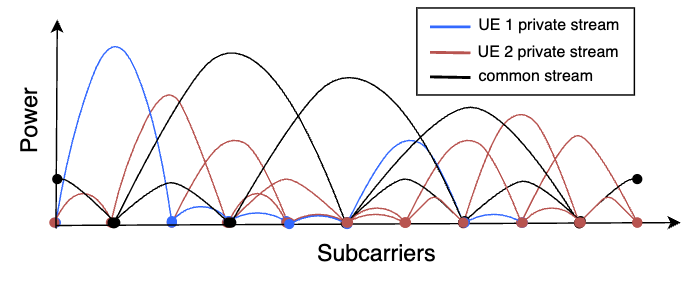} 
    \label{fig:ofdmRSMAscheme}} \vspace{-0.2cm}
    \caption{\ac{RSMA} schemes in different representations, (a) rate-splitting with multiple antennas at BS, (b) rate-splitting in time-frequency domain using OFDM waveform.}
    \label{fig:generalFigure} 
    \vspace{-0.4cm}
\end{figure}

In \cite{Sahin2023OFDMRSMA}, the authors show the advantage of \ac{RSMA} in the \ac{SISO} system signalling over the \ac{OFDM} waveform for the first time. The \ac{OFDM}-\ac{RSMA} scheme is proposed to provide robustness against \ac{ICI}, which allows the interference in the \ac{OFDM} waveform to be partially decoded and partially treated as noise. In this paper, further performance improvement is provided by leveraging multi-numerology concept into the \ac{OFDM}-\ac{RSMA} structure. The proposed multi-numerology \ac{OFDM}-\ac{RSMA} introduces extra degree of freedom to cope with the interference that breaks the orthogonality of \ac{OFDM} subcarriers. Even in SISO system which is not the main attention of prevailing \ac{RSMA} based studies, the proposed multi-numerology \ac{OFDM}-\ac{RSMA} scheme provides the highest achievable sum-rate compared to conventional \ac{OFDMA} and \ac{OFDM}-\ac{NOMA} schemes by keeping fair transmission rates among users at the same time. The main contributions of the paper are as follows:
\begin{itemize}
\item We construct the analogy between \ac{RSMA} in \ac{MISO}-\ac{BC} and multi-numerology \ac{OFDM}-\ac{RSMA} in \ac{SISO} providing extra degree of freedom  to overcome the \ac{ICI}. Common stream's beam of \ac{RSMA} in \ac{MISO}-\ac{BC} is designed to cover both users allowing interference to be partially decoded and partially treated as noise, whereas beams of private streams are directed to the intended users. For the proposed multi-numerology \ac{OFDM}-\ac{RSMA} structure, common stream carrying data of both users also allows interference to partially decoded and partially treated as noise by having larger \ac{SCS} than private streams transmitted in the same time-slot. As it can be seen in Fig. \ref{fig:ofdmRSMAscheme} private streams are designed to be orthogonal and user specific similar to \ac{RSMA} in \ac{MISO}-\ac{BC}. 

\item The system model of the \ac{OFDM}-\ac{NOMA} is mathematically formulated performing the \ac{SIC} at the \ac{OFDM} symbol level rather than the \ac{OFDM} subcarrier level tackled so far in previous studies \cite{Kebede2022_MCNOMA,cejudo2022_NOMAmc,zhu2017_NOMAmc,Lei2016_NOMAmc,salaun2020_NOMApa}. Aforementioned studies assume that the \ac{OFDM}-\ac{NOMA} system is able to both allocate power into subcarriers one by one and change the decoding order at the receiver depending on the power levels of subcarriers which can be fluctuating due to frequency selectivity. In practical wireless systems, the coded-\ac{OFDM} is commonly implemented where the \ac{LLR} of transmitted bits on subcarriers are calculated and then sent to the decoder in a block manner \cite{5G_Dahlman_2021}. This phenomena makes the assumption of subcarrier-dependent decoding order impractical for the \ac{OFDM}-\ac{NOMA} architecture, whereas symbol level \ac{SIC} is performed in this paper to have practical and fair comparison against \ac{OFDM}-\ac{RSMA} and \ac{OFDMA}. 

\item Rate-\ac{AWMMSE} transformations followed by the alternating optimization technique, which is used for \ac{RSMA} in \ac{MISO}-\ac{BC} system \cite{joudeh_2016_SRmax_RSMA}, are adapted to find the optimal subcarrier and power allocation for the proposed multi-numerology \ac{OFDM}-\ac{RSMA} to maximize the sum-rate under the \ac{ICI}. The optimization problem is formulated in a way that the proposed scheme is suitable to perform \ac{OFDM} symbol level transmitter and receiver processing.

\item We perform Monte-Carlo simulations to validate the achievable sum-rate gain achieved by the proposed multi-numerology \ac{OFDM}-\ac{RSMA} compared to \ac{OFDM}-\ac{NOMA} and \ac{OFDMA}. The impact of different system parameters, such as \ac{SCS}, and the channel conditions on the performance of \ac{OFDM}-\ac{RSMA} are also investigated. We also verify that the proposed multi-numerology \ac{OFDM}-\ac{RSMA} reaches more fair rate distribution among users compared to \ac{OFDM}-\ac{NOMA}.
\end{itemize}

The remainder of this paper is organized as follows. Section \ref{sec:SystemModel} describes the utilized channel model and basic \ac{OFDM} transmission. The proposed \ac{OFDM}-\ac{RSMA} method is introduced in Section \ref{sec:OFDMRSMAtransmission}. Section \ref{sec:SolveOFDMRSMA} gives the formulated optimization problem  for the \ac{OFDM}-\ac{RSMA} framework and the proposed algorithm to solbe it. Section \ref{sec:OFDM-NOMA} gives the problem formulation for the \ac{OFDM}-\ac{NOMA} for comparison. Numerical results are presented in Section \ref{sec:Results}. The conclusion of the study is drawn in \ref{sec:Conclusion}.

\textit{Notation:} Lower-case bold face variables indicate vectors, and upper-case bold face variables indicate matrices. We use $\mathbb{C}$ the set of complex numbers, $\lfloor \cdot \rfloor$ to denote floor operation, $\mathbf{I}_N$ to denote the $N\times N$ identity matrix, $\mathbf{0}$ to denote zero vector with appropriate size, $(\cdot)^T$ to denote transpose, $(\cdot)^{\ast}$ to denote complex conjugate, $\Re(\cdot)$ to denote real part of the complex value, $(\cdot)^H$ to denote conjugate transpose, $\mathbb{E}[\cdot]$ to denote expectation. The Euclidean norm, or $\ell_2$-norm, of a vector $x \in \mathbf{R}^n$ is denoted as $\Vert x \Vert_2$ and the \textbf{Frobenius norm} of a matrix $X \in \mathbf{R}^{m \times n}$ is given by $\Vert A \Vert_F$. $\mathbf{A} \odot \mathbf{B}$ and $\mathbf{A} \oslash \mathbf{B}$ correspond to Hadamard multiplication and division of matrices $\mathbf{A}$ and $\mathbf{B}$; $\mathbf{A}^{o2}$ denote Hadamard power of matrix $\mathbf{A}$ by two, $\diag{\mathbf{v}}$ returns a square diagonal matrix with the elements of vector $\mathbf{v}$ on the main diagonal, $\diag{\mathbf{M}}$ returns the elements on the main diagonal of matrix $\mathbf{M}$ in a vector, $\mathcal{C}\mathcal{N}(\mu,\sigma^2)$ represents complex Gaussian random vectors with mean $\mu$ and variance $\sigma^2$. The notation $(\mathbf{A})_n$ is the $n$th diagonal element of the matrix $\mathbf{A}$. Element wise absolute value of vector $\mathbf{a}$ is denoted as $\vert \mathbf{a} \vert$, where $\vert \mathbf{a} \vert = (\vert a_0 \vert, \ldots, \vert a_N \vert  )^T$. The notation $m_{ij}$ is the value located in the $i$th row and the $j$th column of the matrix $\mathbf{M}$ and $\mathbf{e}_i$ denotes the $i$th standard unit basis vector of for $\mathbb{R}^N$. 

\vspace{-0.35cm}
\section{System Model} \label{sec:SystemModel}

We consider the system model with single-antenna transmitter and $K$ single-antenna receivers, indexed by $\mathcal{K} = \{1,2,\ldots,K\}$. In conventional orthogonal transmission, the transmitter performs \ac{OFDMA} to serve the $K$ users in the allocated time-frequency resource. Channel model and signalling scheme are explained in the following. 

\vspace{-0.35cm}
\subsection{Channel Model}

Throughout the study, time-varying frequency selective fading channels have been used, which are named as doubly dispersive or doubly selective. It produces time and frequency shifts of the transmitted signal due to multipath propagation and Doppler effect. The channel model includes complex channel gain, Doppler shift and delay for every path. Therefore, the propagation channel in the time-delay domain, $c(t,\tau)$, can be modeled as follows \cite{hlawatsch_matz_2011}:
\vspace{-0.2cm}
\begin{IEEEeqnarray}{rCl}
    c(t,\tau)  = \sum_{l=1}^L \alpha_l e^{j2\pi\nu_l t}  \delta(\tau-\tau_l),
\label{eq:channelModel} 
\vspace{-0.15cm}
\end{IEEEeqnarray}
where $\alpha_l$, $\tau_l$, and $\nu_l$ denote the complex attenuation factor, time delay, and Doppler frequency shift associated with the $l^{\text{th}}$ path where $l \in \left \lbrace 1,2,\ldots, L \right \rbrace$. Let $N$ and $C$ be the subcarrier number and \ac{CP} length of the \ac{OFDM} waveform, respectively. It is assumed that \ac{CP} length is larger than the maximum delay spread to ensure \ac{ISI} free transmission. The relation of (\ref{eq:channelModel}) with the $k$th user's time domain channel matrix $\mathbf{H}_k$ can be represented as follows: 
\vspace{-0.15cm}
\begin{IEEEeqnarray}{rCl}
\mathbf{H}_k  = \sum_{l=1}^L \alpha_l \mathbf{\Pi}^{n_{\tau_l}}  \mathbf{\Delta}(\nu_l),
\label{eq:channelModelMatrix} 
\vspace{-0.1cm}
\end{IEEEeqnarray} 
where the delay matrix $\mathbf{\Pi}^{n_{\tau_l}} \in \mathbb{C}^{(N+C) \times (N+C)}$ is the forward cyclic shifted permutation matrix according to the delay of the $l$th path. The delay matrix $\mathbf{\Pi}^{n_{\tau_l}}$ can be expressed as follows:
\vspace{-0.15cm}
\begin{IEEEeqnarray}{rCl}
    \mathbf{\Pi} = \left[\begin{array}{ccccc}
        0 & 0 & \cdots & 0 & 1 \\
        1 & 0 & \cdots & 0 & 0 \\
        0 & 1 & \cdots & 0 & 0 \\
        \vdots & \vdots & \ddots & \vdots & \vdots \\
        0 & 0 & \cdots & 1 & 0 \\
\end{array}\right], \quad \text{and} \quad n_{\tau_l} = \left \lfloor \frac{\tau_l}{F_s} \right \rfloor, 
\vspace{-0.1cm}
\end{IEEEeqnarray}
where $F_s$ is the sampling frequency in the system. The Doppler shift matrix for the $l$th path, $\mathbf{\Delta}(\nu_l) \in \mathbb{C}^{(N+C) \times (N+C)} $, can be written as follows: 
\vspace{-0.15cm}
\begin{IEEEeqnarray}{rCl}
    \mathbf{\Delta}(\nu_l) = \diag{\left[e^{\frac{j2\pi\nu_l}{F_s}}, e^{\frac{j2\pi\nu_l 2}{F_s}}, \cdots, 
    e^{\frac{j2\pi\nu_l (N+C)}{F_s}}\right]}.
    \vspace{-0.15cm}
\end{IEEEeqnarray}
Let $\Tilde{\mathbf{H}}_k$ be the complete \ac{CFR} matrix of the $k$th user's channel which is obtained as follows: 
\vspace{-0.6cm}
\begin{IEEEeqnarray}{rCl}
    \Tilde{\mathbf{H}}_k &=& \mathbf{F} \mathbf{B} \mathbf{H}_k \mathbf{A} \mathbf{F}^H,
    \label{eq:channelCFR}
    \vspace{-0.2cm}
\end{IEEEeqnarray}
where $\mathbf{F} \in \mathbb{C}^{N \times N}$ is the $N$-point \ac{FFT} matrix, the \ac{CP}-addition matrix $\mathbf{A} \in \mathbb{N}^{(N+C) \times N}$, and the \ac{CP}-removal matrix $\mathbf{B}_c \in \mathbb{N}^{N \times (N+C)}$, are defined as follows: 
\vspace{-0.1cm}
\begin{IEEEeqnarray}{rCl} 
	\mathbf{A} &=& \begin{bmatrix}
	\mathbf{0}_{C \times (N-C)} & \mathbf{I}_{C} \\ 
	\multicolumn{2}{c}{\mathbf{I}_N} 
	\end{bmatrix}, \; \mathbf{B} = \begin{bmatrix}
	\mathbf{0}_{N \times C} & \mathbf{I}_N 
	\end{bmatrix}.
	\nonumber
	\IEEEeqnarraynumspace
 \vspace{-0.1cm}
\end{IEEEeqnarray}
The diagonal components of \ac{CFR} shown in (\ref{eq:channelCFR}) are the $k$th user's channel coefficients scaling the subcarrier in interest, shown with a vector $\mathbf{h}_k \in \mathbb{C}^{N\times 1}$ where
\vspace{-0.1cm}
\begin{IEEEeqnarray}{rCl}
    \mathbf{h}_k &=& \diag{\Tilde{\mathbf{H}}_k},
    \vspace{-0.1cm}
\end{IEEEeqnarray}
where $k \in \mathcal{K} = \{1,2,\ldots,K\}$. It should be noted that off-diagonal components of \ac{CFR} represent the \ac{ICI} due to time variations of the propagation channel. 

\begin{figure}[t]
\vspace{-0.2cm}
    \centering
    \subfloat[]{
    \includegraphics[width=1\linewidth]{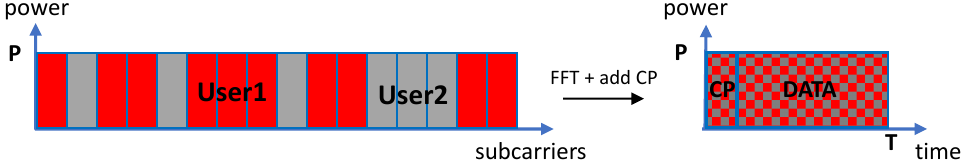}
    \label{fig:OFDMAschemes1}}
    \vspace{-0.3cm} \hfill
    \subfloat[]{
    \includegraphics[width=1\linewidth]{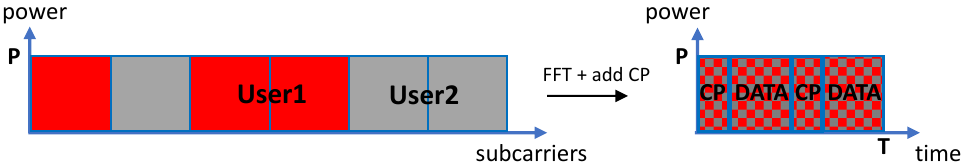}\label{fig:OFDMAschemes2}}
    \vspace{-0.3cm}\hfill
    \subfloat[]{
    \includegraphics[width=1\linewidth]{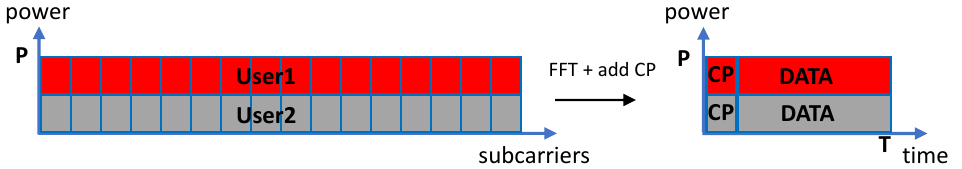}\label{fig:NOMAschemes1}}
    \vspace{-0.3cm}\hfill
    \subfloat[]{
    \includegraphics[width=1\linewidth]{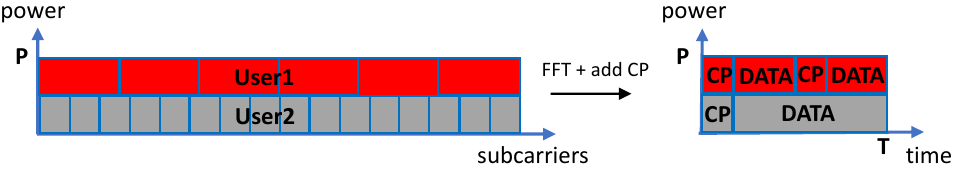}\label{fig:NOMAschemes2}}
    \vspace{-0.3cm}\hfill
    \subfloat[]{
    \includegraphics[width=1\linewidth]{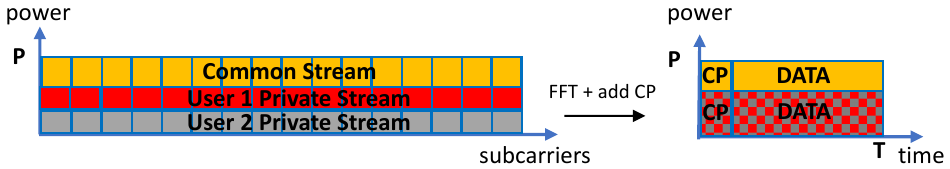}\label{fig:RSMAschemes1}}
    \vspace{-0.3cm}\hfill
    \subfloat[]{
    \includegraphics[width=1\linewidth]{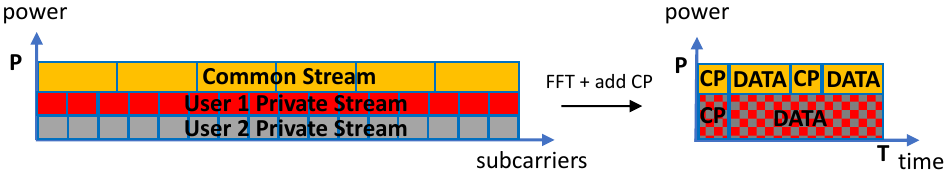}\label{fig:RSMAschemes2}} \vspace{-0.2cm}
    \caption{OFDM based multiple access schemes, (a) OFDMA with narrow \acp{SCS}, (b) OFDMA with wide \acp{SCS}, (c) \ac{OFDM}-\ac{NOMA} with two users having same \acp{SCS}, (d) \ac{OFDM}-\ac{NOMA} with two users having different SCSs, (e) \ac{OFDM}-\ac{RSMA} with two users having same \ac{SCS} for all three streams, (f) \ac{OFDM}-\ac{RSMA} with two users having different \ac{SCS} between common stream and private streams.}
    \label{fig:rsmaGeneralization} \vspace{-0.4cm}
\end{figure}

\vspace{-0.2cm}
\subsection{OFDM Transmission for $K$ Users}

Throughout the study, it is considered that each user can have different \ac{SCS}. \ac{CP} duration of \ac{OFDM} symbols is $C$ for every user to provide equal robustness against delay spread. Subscript notations are added to FFT, CP addition and removal matrices to differentiate between users. The stream of $k$th user with the \ac{OFDM} modulation can be expressed as follows: 
\vspace{-0.15cm}
\begin{IEEEeqnarray}{rCl}
    \mathbf{x}_k = \mathbf{A}_k \mathbf{F}_k^H \diag{\mathbf{p}_k}\mathbf{d}_k, \quad k \in \mathcal{K}, 
    \label{eq:NOMAsignal}
    \vspace{-0.15cm}
\end{IEEEeqnarray}
where $\mathbf{A}_k, \mathbf{F}_k, \mathbf{p}_k$, and $\mathbf{d}_k$ denote \ac{CP}-addition matrix, \ac{FFT} matrix, precoding vector, and the communication signal for the $k$th user, respectively. The time-domain received signal, $\mathbf{y}_k$, after passing through the $k$th user channel $\mathbf{H}_k$, can be written as follows:
\vspace{-0.15cm}
\begin{IEEEeqnarray}{rCl} 
	\mathbf{y}_{k} &=& \mathbf{H}_{k} \sum_{k=1}^K \mathbf{x}_{k} + \mathbf{n}_k.
	\label{eq:NOMAtransmittedAfterChannel}
 \vspace{-0.15cm}
\end{IEEEeqnarray}
where the vector $\mathbf{n}_k \in \mathbb{C}^{(C + N_p) \times 1}$ is the \ac{AWGN} with $n_{k_i} \sim \mathcal{CN}(0,\sigma^2)$ where $n_{k_i} \in \mathbf{n}_k$ and $\sigma^2$ is the power of \ac{AWGN}.

\section{The Proposed \ac{OFDM}-\ac{RSMA} Transmission Scheme}
\label{sec:OFDMRSMAtransmission}

\begin{figure*}[t]   
\vspace{-0.1cm}
    \centering
\includegraphics[width=1\linewidth]{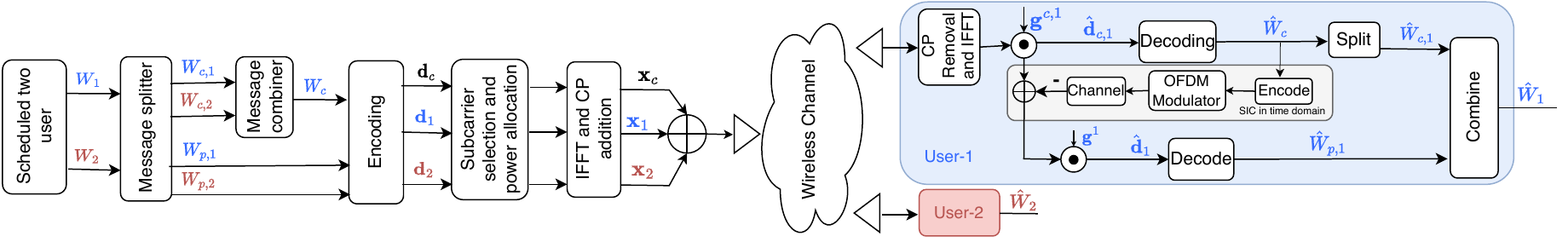}
\vspace{-0.4cm}
    \caption{Proposed \ac{OFDM}-\ac{RSMA} transceiver structure.} \label{fig:ofdmRSMAtransceiver}
    \vspace{-0.5cm}
\end{figure*}

In this section, the proposed \ac{OFDM}-\ac{RSMA} scheme is introduced where additional data stream modulated via \ac{OFDM} waveform is integrated into users' dedicated signals utilizing \ac{RS} based allocation to provide robustness against interference sources, {\it i.e}, \ac{ICI}, \ac{INI}, and \ac{MUI}. The message intended for user-$k$, $W_k$, is split into common and private parts, which are denoted as $W_{c,k}$ and $W_{p,k}$, $\forall k \in \mathcal{K}$. The common parts of messages for all users are combined into a common message $W_{c}$ which is encoded into common stream, $\mathbf{d}_c$. Private part of messages, $W_{p,k}$, are independently encoded into $\mathbf{d}_{k}, \forall k \in \mathcal{K}$. Fig. \ref{fig:ofdmRSMAtransceiver} demonstrates the proposed \ac{RS}-based transmission scheme using the  \ac{OFDM} waveform in two users scenario.
The private parts of the split messages of the users are carried on $N_{p}$ subcarriers with $\Delta f_{p}$ \ac{SCS}. The \ac{SCS} and number of subcarriers for the combined common message of the users are denoted as $\Delta f_{c}$ and $N_{c}$, respectively. The subcarrier sets are defined for common and private streams as $\mathcal{N}_c = \{1,\ldots, N_c\}$ and $\mathcal{N}_p = \{1,\ldots, N_p\}$, respectively. The transmitter uses a cyclic prefix (CP) of length $C$ for private and common streams. For a $K$-user system, the transmitted private stream $\mathbf{x}_k \in \mathbb{C}^{(C + N_p)\times 1}$ for user-$k$, and common stream $\mathbf{x}_{c,m} \in \mathbb{C}^{(C + N_p)\times 1}$, are expressed as follows:
\vspace{-0.2cm}
\begin{IEEEeqnarray}{rCl} 
    \mathbf{x}_{k} &=& \mathbf{A}_{p}  \mathbf{F}_{p}^H \diag{\mathbf{p}_{k}} \mathbf{d}_{k},    \IEEEyesnumber  \label{eq:rsma_streams} \IEEEyessubnumber \label{eq:privateStream} \\  
    \mathbf{x}_{c,m} &=& \mathbf{A}_{c,m}  \mathbf{F}_c^H \diag{\mathbf{p}_{c,m}} \mathbf{d}_{c,m}, \quad \forall m \in \mathcal{M} \IEEEyessubnumber, \label{eq:commonStream} 
    \vspace{-0.1cm}
\end{IEEEeqnarray}
where the set $\mathcal{M} = \left\{1,\ldots, M \right\}$ defines the number of common stream's OFDM symbol that can be confined in the duration of private stream with $M=\left\lfloor\frac{C+N_p}{C+N_c}\right\rfloor$. 
The \ac{CP}-addition matrices for private and common streams, $\mathbf{A}_p \in \mathbb{N}^{(C+N_p) \times N_p}$ and $\mathbf{A}_{c,m} \in \mathbb{N}^{(C+N_p) \times N_c}$, are defined as follows: 
\begin{IEEEeqnarray}{rCl} 
    \mathbf{A}_p &=& \begin{bmatrix}
    \mathbf{0}_{C \times (N_p-C)} & \mathbf{I}_{C} \\ 
    \multicolumn{2}{c}{\mathbf{I}_{N_p}} 
    \end{bmatrix}, \mathbf{A}_{c,m} = \begin{bmatrix}
     \multicolumn{2}{c}{\mathbf{0}_{(m-1) (C+N_c)}} \\ 
    \mathbf{0}_{C \times (N_c-C)} & \mathbf{I}_{C} \\ 
    \multicolumn{2}{c}{\mathbf{I}_{N_c}} \\ \multicolumn{2}{c}{\mathbf{0}_{(N_p+C-m(N_c+C)) \times N_c}}
    \end{bmatrix}, \IEEEeqnarraynumspace
    \nonumber
\end{IEEEeqnarray}
where $\mathbf{F}_p \in \mathbb{C}^{N_p \times N_p}$ is the $N_p$-point \ac{FFT} matrix, whereas $\mathbf{F}_c \in \mathbb{C}^{N_c \times N_c}$ denotes the $N_c$-point \ac{FFT} matrix. The communication signals $\mathbf{d}_{k} \in \mathbb{C}^{N_{p} \times 1}$ for $k$th user private stream, and $\mathbf{d}_{c,m} \in \mathbb{C}^{N_c \times 1}$ for $m$th common stream are chosen independently from a Gaussian alphabet for theoretical analysis. It is assumed that the streams have unit power, $\mathbb{E} \{ \tilde{\mathbf{d}}\tilde{\mathbf{d}}^H\} = \mathbf{I}$, where $\tilde{\mathbf{d}} = [\mathbf{d}_{c,1}^T,\ldots, \mathbf{d}_{c,M}^T, \mathbf{d}_1^T, \ldots, \mathbf{d}_K^T]^T $. The vectors $\mathbf{p}_{k} \in \mathbb{C}^{N_p \times 1}$, and $\mathbf{p}_{c,m} \in \mathbb{C}^{N_c \times 1}$ are amplitude values allocated to private and common streams, respectively. The matrix $\mathbf{P} = [\mathbf{p}_1,\ldots,\mathbf{p}_K]$ is defined as the collection of all amplitude vectors of private users, $\mathbf{p}_k, \forall k \in \mathcal{K}$. Similarly $\mathbf{P}_c = [\mathbf{p}_{c,1}^T,\ldots,\mathbf{p}_{c,M}^T]^T$ corresponds to the collection of all amplitude vectors of common streams, $\mathbf{p}_{c,m}, \forall m \in \mathcal{M}$. 

The time-domain received signal,  $\mathbf{y}_k \in \mathbb{C}^{(C + N_p) \times 1}$, after passing through the $k$th user channel $\mathbf{H}_k \in \mathbb{C}^{(C + N_p) \times (C + N_p)}$, can be written as follows:
\vspace{-0.1cm}
\begin{IEEEeqnarray}{rCl} 
    \mathbf{y}_{k} &=& \mathbf{H}_{k} \left( \sum_{m=1}^M \mathbf{x}_{c,m} + \sum_{k=1}^K \mathbf{x}_{k} \right) + \mathbf{n}_k.
    \label{eq:transmittedAfterChannel}
    \vspace{-0.1cm}
\end{IEEEeqnarray}
At the $k$th user's receiver, the superimposed signal $\mathbf{y}_k$ is processed with CP removal matrix and FFT operation to convert into frequency domain for equalization and demodulation. Note that the \ac{SCS} of common stream may be larger than the \ac{SCS} of common streams from the design perspective, therefore, common and private streams are treated separately. To process the common stream, received signals of users are obtained after passing through the common stream \ac{CP} removal matrix and \ac{FFT} matrix whose size is suitable for the length of common stream. The received frequency domain signal at the $k$th user's $m$th OFDM symbol, $\mathbf{r}_{c,k,m}$ can be expressed mathematically as follows:
\vspace{-0.15cm}
\begin{IEEEeqnarray}{rCl} 
    \mathbf{r}_{c,k,m} &=& \mathbf{F}_c \mathbf{B}_{c,m}\mathbf{y}_{k},	\label{eq:commonSequence_received}
    \vspace{-0.15cm}
\end{IEEEeqnarray}
$\mathbf{B}_{c,m} = [\mathbf{0}_{N_c \times (C+(m-1)(C+N_c))}, \mathbf{I}_{N_c}, \mathbf{0}_{N_c \times (C+N_p-m(N_c+C))}] $ is the common stream \ac{CP} removal matrix. The average received power at the $n$th subcarrier ($n \in \mathcal{N}_c$) of $\mathbf{r}_{c,k,m}$ for the certain channel state, $T_{c,k,m,n} \triangleq \mathsf{E} \left\{ \vert {(\mathbf{r}_{c,k,m})}_n \vert^2 \right\}$, is written as:
\vspace{-0.1cm}
\begin{IEEEeqnarray}{rCl} \label{eq:power_common} 
    T_{c,k,m,n} & = &  \left \vert s^{c,k,m,n}_{n,n} \right \vert^2 + \underbrace{ \sum_{j=1}^{N_c} \left \vert \bar{s}^{c,k,m,n}_{n,j} \right \vert^2 + \sum_{u = 1}^K \sum_{j=1}^{N_p} \left \vert s^{u,m}_{n,j} \right \vert^2 + \sigma^2}_{I_{c,k,m,n}}, \nonumber \\ 
    \vspace{-0.15cm}
\end{IEEEeqnarray}
where  
\vspace{-0.15cm}
\begin{IEEEeqnarray}{rCl} 	
    \mathbf{S}^{c,k,m,n} &=& \mathbf{F}_c\mathbf{B}_{c,m} \mathbf{H}_k \mathbf{A}_{c,m} \mathbf{F}_c^H \diag{p_{c,m,n}\mathbf{e}_n}, \;\forall n \in \mathcal{N}_c,  \nonumber \IEEEeqnarraynumspace \\ 
    \bar{\mathbf{S}}^{c,k,m,n} &=& \mathbf{F}_c\mathbf{B}_{c,m} \mathbf{H}_k \mathbf{A}_{c,m} \mathbf{F}_c^H \diag{\bar{\mathbf{p}}_{c,m,n}}, \; \forall n \in \mathcal{N}_c, \nonumber\\
    \mathbf{S}^{u,m} &=& \mathbf{F}_c\mathbf{B}_{c,m} \mathbf{H}_k \mathbf{A}_p \mathbf{F}_p^H \diag{\mathbf{p}_{u}}, \quad \forall u \in \mathcal{K}, \nonumber \IEEEeqnarraynumspace
    \vspace{-0.15cm}
\end{IEEEeqnarray}
where $\forall k \in \mathcal{K}$, $\forall m \in \mathcal{M}$ and  $\bar{\mathbf{p}}_{c,m,n}$ denotes that $n$th subcarrier is forced not to carry any energy for the common stream's $m$th \ac{OFDM} symbol, i.e.,  $\bar{\mathbf{p}}_{c,m,n} = [p_{c,m,1},\ldots,p_{c,m,n-1},0,p_{c,m,n+1},\ldots,p_{c,m,N_c}]^T$. The $k$th user's receiver demodulates the common stream, then reconstructs and subtracts from the received signal. With the removal of common stream successfully, \ac{FFT} matrix succeeding \ac{CP} removal matrix is applied to the received signal in order to demodulate corresponding private stream as follows:
\vspace{-0.15cm}
\begin{IEEEeqnarray}{rCl} 
    \mathbf{r}_{k} &=& \mathbf{F}_p \mathbf{B}_p\left(\mathbf{y}_k - \mathbf{H}_k \sum_{m=1}^M \mathbf{A}_{c,m}  \mathbf{F}_c^H \diag{\mathbf{p}_{c,m}} \mathbf{d}_{c,m}\right) \nonumber \\ &=& \mathbf{F}_p \mathbf{B}_p \left( \mathbf{H}_k \sum^K_{k=1} \mathbf{x}_k + \mathbf{n}_k\right), \IEEEeqnarraynumspace	\label{eq:privateSequence_received}
    \vspace{-0.15cm}
\end{IEEEeqnarray}
where $\mathbf{B}_p = [\mathbf{0}_{N_p \times C}, \mathbf{I}_{N_p}] $ is the private stream \ac{CP} removal matrix. The average received power at the $q$th subcarrier of $\mathbf{r}_{k}$, $T_{k,q} \triangleq \mathsf{E} \left\{ \left\vert {(\mathbf{r}_{k})}_q \right \vert^2 \right\}$, can be written as:
\vspace{-0.15cm}
\begin{IEEEeqnarray}{rCl} 
    T_{k,q} & = & \left \vert v^{k,q}_{q,q} \right \vert^2 + \underbrace{ \sum_{j=1}^{N_p} \left \vert \bar{v}^{k,q}_{q,j} \right \vert^2 + \sum_{\substack{i=1 \\ i \neq k}}^{K} \sum_{j=1}^{N_p} \left \vert w^i_{q,j}  \right \vert^2 + \sigma^2}_{I_{k,q}}, \IEEEeqnarraynumspace \label{eq:power_private}
    \vspace{-0.2cm}
\end{IEEEeqnarray}	
where
\vspace{-0.2cm}
\begin{IEEEeqnarray}{rCl} 	
    \mathbf{V}^{k,q} &=& \mathbf{F}_p \mathbf{B}_p \mathbf{H}_k \mathbf{A}_p \mathbf{F}_p^H \diag{p_{k,q}\mathbf{e}_q}, \; \forall k \in \mathcal{K}, \; \forall q \in \mathcal{N}_p,  \nonumber \IEEEeqnarraynumspace \\
    \bar{\mathbf{V}}^{k,q} &=& \mathbf{F}_p \mathbf{B}_p \mathbf{H}_k \mathbf{A}_p \mathbf{F}_p^H \diag{\bar{\mathbf{p}}_{k,q}}, \; \forall k \in \mathcal{K}, \; \forall q \in \mathcal{N}_p, \nonumber \IEEEeqnarraynumspace \\
    \mathbf{W}^{i} &=& \mathbf{F}_p \mathbf{B}_p \mathbf{H}_k \mathbf{A}_p \mathbf{F}_p^H \diag{\mathbf{p}_i}, \; \forall i \in {\mathcal{K} \backslash k}, \nonumber \IEEEeqnarraynumspace
    \vspace{-0.1cm}
\end{IEEEeqnarray}
where $\bar{\mathbf{p}}_{k,q}$ denotes that $q$th subcarrier is forced not to carry any energy for private and common streams, i.e.,  $\bar{\mathbf{p}}_{k,q} = [p_{q,1},\ldots,p_{k,q-1},0,p_{k,q+1},\ldots,p_{k,N_p}]^T$. 

By using (\ref{eq:power_common}) and (\ref{eq:power_private}) SINRs of the common and private streams for a given channel state can be stated as follows: 
\vspace{-0.1cm}
\begin{IEEEeqnarray}{rCl}
   \gamma_{c,k,m,n} &\triangleq& \left \vert s^{c,k,m,n}_{n,n} \right \vert^2 I^{-1}_{c,k,m,n} \quad \text{and} \quad \gamma_{k,q} \triangleq \left \vert v^{k,q}_{q,q} \right \vert^2  I^{-1}_{k,q}. \IEEEeqnarraynumspace \label{eq:SINRs}
   \vspace{-0.1cm}
\end{IEEEeqnarray}
The achievable rates for common stream and private streams corresponding to $k$th user in the corresponding subcarriers can be written as follows:
\vspace{-0.1cm}
\begin{IEEEeqnarray}{rCl}
   R_{c,k,m,n} &=& \log_2(1+\gamma_{c,k,m,n}), \quad R_{k,q} = \log_2(1+\gamma_{k,q}), \IEEEeqnarraynumspace \label{eq:achievableRates}
   \vspace{-0.1cm}
\end{IEEEeqnarray}
and the achievable rates for an OFDM symbol can be written as follows: 
\vspace{-0.1cm}
\begin{IEEEeqnarray}{rCl}
   R_{c,k} &=& \sum_{m=1}^M \sum_{n=1}^{N_c} R_{c,k,m,n}  \; \text{bit/s/Hz,} \;  \; R_{k} = \sum_{q=1}^{N_q} R_{k,q} \; \text{bit/s/Hz}.  \label{eq:achievableRatesOFDM} \IEEEeqnarraynumspace
   \vspace{-0.1cm}
\end{IEEEeqnarray}
In the next section, problem formulation to maximize the sum data-rate of the proposed OFDM-RSMA system is provided. Alternation optimization is followed by \ac{WMMSE}-based transformation to solve the optimization problem. 

\vspace{-0.1cm}
\section{Problem Formulation and Proposed Algorithm for \ac{OFDM}-\ac{RSMA}} \label{sec:SolveOFDMRSMA}
\vspace{-0.1cm}

The optimization problem for achievable rate maximization using \ac{RSMA} with \ac{OFDM} waveform can be formulated as follows:
\vspace{-0.1cm}
\begin{maxi}
{\mathbf{R}_c,\mathbf{P}_c,\mathbf{P}}{ \sum_{m=1}^M \sum_{n=1}^{N_c} R_{c,k,m,n} + \sum^K_{k=1} \sum_{q=1}^{N_p} R_{k,q}}
{\label{eq:optProblem}}{}
\addConstraint{\sum_{m=1}^M\sum_{n=1}^{N_c} R_{c,k,m,n}}{\geq \sum_{m=1}^M \sum_{n=1}^{N_c} R_{c,m,n}},
\addConstraint{\Vert \mathbf{P} \Vert^2_F + \Vert \mathbf{P}_c \Vert_F^2}{\leq P_t},
\addConstraint{\sum_{n=1}^{N_c} C_{k,n} + \sum_{q=1}^{N_p} R_{k,q}}{\geq R_k^{\text{min}}, \forall k \in \mathcal{K}},
\vspace{-0.1cm}
\end{maxi}
where the common rate at the $n$th subcarrier $R_{c,m,n}$ is shared among users such that $C_{k,m,n}$ is the $k$th user's portion of the common rate with $R_{c,m,n} = \sum_{k=1}^K C_{k,m,n}$ and $\mathbf{R}_c$ is the collection of $R_{c,k,m,n}, \forall k \in \mathcal{K}, \forall m \in \mathcal{M}, \forall n \in \mathcal{N}_c$ . 

Let $\hat{\mathbf{d}}_{c,k,m} = \mathbf{g}^{c,k,m} \odot \mathbf{F}_c \mathbf{B}_{c,m} \mathbf{y}_k$ be $k$th user's estimate of $\mathbf{d}_{c,m}$, where $\mathbf{g}^{c,k,m} \in \mathbb{C}^{N_c \times 1}$ is the one tap-equalizer vector for $m$th OFDM symbol. After successfully removing the common stream, the estimate of ${\mathbf{d}}_{k}$ can be obtained as follows: 
\begin{IEEEeqnarray}{rCl}
    \hat{\mathbf{d}}_{k} &=& \mathbf{g}^{k} \odot \mathbf{F}_{p} \mathbf{B}_{p} \left(\mathbf{y}_k - \mathbf{H}_k \sum_{m=1}^M \mathbf{A}_{c,m}  \mathbf{F}_c^H \diag{\mathbf{p}_{c,m}} \hat{\mathbf{d}}_{c,k,m}\right), \IEEEeqnarraynumspace \nonumber
\end{IEEEeqnarray}
where $\mathbf{g}^{k}$ is the corresponding one-tap equalizer vector. At the output of the $k$th receiver, the common and private \acp{MSE} for $n$th subcarrier in the $m$th common stream and $q$th subcarrier in the private stream are defined as $\varepsilon_{c,k,m,n} \triangleq \mathsf{E}\left\{ \left \vert \hat{d}_{c,k,m,n} - d_{c,m,n} \right \vert^2 \right\}$
and $\varepsilon_{k,q} \triangleq \mathsf{E}\left\{ \left \vert \hat{d}_{k,q} - d_{k,q} \right \vert^2 \right\}$ respectively, which can be written as follows under the assumptions that common stream has been decoded and removed perfectly:
\vspace{-0.15cm}
\begin{IEEEeqnarray}{rCl} 
    \varepsilon_{c,k,m,n} &=& \left \vert g^{c,k,m}_n \right \vert^2 T_{c,k,m,n} - 2\mathfrak{R}\left\{ g^{c,k,m}_n s^{c,k,m,n}_{n,n} \right\} + 1, \IEEEyesnumber \label{eq:MSE_both} \IEEEyessubnumber \label{eq:MSE_common} \IEEEeqnarraynumspace \\
    \varepsilon_{k,q} &=& \left \vert g^{k}_q \right \vert^2 T_{k,q} - 2\mathfrak{R}\left\{ g^k_q v^{k,q}_{q,q} \right\} + 1. \IEEEyessubnumber \label{eq:MSE_private}
    \vspace{-0.1cm}
\end{IEEEeqnarray}
Optimum \ac{MMSE} equalizers for corresponding subcarriers can be found by solving $\frac{\partial \varepsilon_{c,k,m,n}}{\partial g^{c,k,m}_n} = 0$ for $n \in \mathcal{N}_c$, and $\frac{\partial \varepsilon_{k,q}}{\partial g^k_q} = 0$ for $q \in \mathcal{N}_p$, 
\vspace{-0.1cm}
\begin{IEEEeqnarray}{rCl} 
    \left({g^{c,k,m}_n}\right)^{\text{MMSE}} &=& T_{c,k,m,n}^{-1} \left(s^{c,k,m,n}_{n,n}\right)^H,   \IEEEyesnumber \label{eq:Optimum_equalizer} \IEEEyessubnumber \label{eq:Optimum_equalizer_1} \\ \left({g^{k}_q}\right)^{\text{MMSE}} &=& T_{k,q}^{-1} \left({v^{k,q}_{q,q}}\right)^H. \IEEEyessubnumber \label{eq:Optimum_equalizer_2}
    \vspace{-0.1cm}
\end{IEEEeqnarray}
After substituting optimum equalizers found in (\ref{eq:Optimum_equalizer}) into equations in (\ref{eq:MSE_both}), \acp{MMSE} can be written as follows: 
\vspace{-0.1cm}
\begin{IEEEeqnarray}{rCl} 
    \varepsilon^{\text{MMSE}}_{c,k,m,n} &\triangleq& \min_{g^{c,k,m}_n} \varepsilon_{c,k,m,n} =  T_{c,k,m,n}^{-1} I_{c,k,m,n}, \IEEEyesnumber \label{eq:Optimum_MMSE} \IEEEyessubnumber \label{eq:Optimum_MMSE_1} \IEEEeqnarraynumspace \\
    \varepsilon_{k,q}^{\text{MMSE}} &\triangleq& \min_{g^{k}_n} \varepsilon_{k,q} =  T_{k,q}^{-1} I_{k,q}. \IEEEyessubnumber \label{eq:Optimum_MMSE_2} \IEEEeqnarraynumspace
    \vspace{-0.1cm}
\end{IEEEeqnarray}
From (\ref{eq:Optimum_MMSE}), \acp{SINR} in (\ref{eq:SINRs}) and achievable rates in (\ref{eq:achievableRates}) for corresponding subcarriers can be rewritten as follows \cite{cioffi_2008_WMMSE}: 
\vspace{-0.1cm}
\begin{IEEEeqnarray}{rCl}
    \gamma_{c,k,m,n} &=& \frac{1-\varepsilon^{\text{MMSE}}_{c,k,m,n}}{\varepsilon^{\text{MMSE}}_{c,k,m,n}}, \quad \text{and} \quad
    \gamma_{k,q} = \frac{1-\varepsilon_{k,q}^{\text{MMSE}}}{\varepsilon_{k,q}^{\text{MMSE}}}, \label{eq:newSINRs} \IEEEeqnarraynumspace \\ 
    R_{c,k,m,n} &=& -\log_2(\varepsilon^{\text{MMSE}}_{c,k,m,n}), \quad 
    R_{k,q} = -\log_2(\varepsilon_{k,q}^{\text{MMSE}}). \IEEEeqnarraynumspace \label{eq:newAchievableRates}
    \vspace{-0.1cm}
\end{IEEEeqnarray}
To solve the optimization problem of (\ref{eq:achievableRatesOFDM}), the \ac{WMMSE} method \cite{cioffi_2008_WMMSE} is adapted to the \ac{OFDM}-\ac{RSMA} structure. Inspired from \cite{joudeh_2016_SRmax_RSMA}, the \ac{AWMMSE} of common and private streams can be written as follows:
\vspace{-0.1cm}
\begin{IEEEeqnarray}{rCl}
    \zeta_{c,k,m,n} &=& u^{c,k,m}_n \varepsilon_{c,k,m,n} - \log_2\left(u^{c,k,m}_n\right), \IEEEyesnumber \label{eq:AWMSEs} \IEEEyessubnumber \label{eq:AWMSEsCommon} \\  
    \zeta_{k,q} &=& u^k_q \varepsilon_{k,q} - \log_2\left(u^k_q\right), \IEEEyessubnumber \label{eq:AWMSEsPrivate}
    \vspace{-0.1cm}
\end{IEEEeqnarray}
where $u^{c,k,m}_n, u^k_q > 0$ are weights associated with the $k$th user's MSEs of common and private stream in the corresponding subcarrier, respectively. To find the rate-\ac{AWMMSE} relationship, (\ref{eq:AWMSEs}) needs to be optimized regarding to equalizers and weights. By solving $\frac{\partial \zeta_{c,k,m,n}}{\partial g^{c,k,m}_n} = 0$ and $\frac{\partial \zeta_{k,q}}{\partial g^k_q} = 0$, the optimum equalizers can be found as 
$\left(g^{c,k,m}_n\right)^{\ast} = \left(g^{c,k,m}_n \right)^{\text{\ac{MMSE}}}$ and 
$\left(g^{k}_q\right)^{\ast} = \left(g^k_q\right)^{\text{\ac{MMSE}}}$. Substituting these optimum equalizers into (\ref{eq:AWMSEs}) yields
\vspace{-0.1cm}
\begin{IEEEeqnarray}{rCl}               
    \zeta_{c,k,m,n}\left(\left(g^{c,k,m}_n\right)^{\ast}\right) &=& u^{c,k,m}_n\varepsilon^{\text{MMSE}}_{c,k,m,n} - \log_2 \left(u^{c,k,m}_n\right), \IEEEyesnumber \label{eq:newAWMSEs} \IEEEyessubnumber \label{eq:newAWMSEs_1} \IEEEeqnarraynumspace \\ 
    \zeta_{k,q}\left(\left(g^k_q\right)^{\ast}\right) &=& u^k_q \varepsilon^{\text{MMSE}}_{k,q} - \log_2\left(u^k_q\right). \IEEEyessubnumber \label{eq:newAWMSEs_2}
    \vspace{-0.1cm}
\end{IEEEeqnarray}
Then, taking derivative  with respect to weight factors and equating to zero as $\frac{\partial \zeta_{c,k,m,n}((g^{c,k,m}_n)^{\ast})}{\partial u^{c,k,m}_n} = 0$ and $\frac{\partial \zeta_{k,q}((g^k_q)^{\ast})}{\partial u^k_q} = 0$, optimum \ac{MMSE} weights can be obtained as follows: 
\vspace{-0.1cm}
\begin{IEEEeqnarray}{rCl}
    \left(u^{c,k,m}_n \right)^{\ast} &=& \left(u^{c,k,m}_n\right)^{\text{MMSE}} \triangleq  \left(\varepsilon^{\text{MMSE}}_{c,k,m,n}\right)^{-1}, \IEEEyesnumber \label{eq:optimumWeights} \IEEEyessubnumber \label{eq:optimumWeights_1} \\ 
    \left(u^k_q\right)^{\ast} &=& \left(u^k_q\right)^{\text{MMSE}} \triangleq  \left(\varepsilon^{\text{MMSE}}_{k,q}\right)^{-1},\IEEEyessubnumber \label{eq:optimumWeights_2}
    \vspace{-0.1cm}
\end{IEEEeqnarray}
where the scaling factor of $(\ln(2))^{-1}$ has been omitted without having an impact. By substituting these into (\ref{eq:newAWMSEs}), the rate-WMMSE relationship is formulated as follows: 
\vspace{-0.1cm}
\begin{IEEEeqnarray}{rCl}
    \zeta^{\text{MMSE}}_{c,k,m,n} &\triangleq& \min_{u^{c,k,m}_n,g^{c,k,m}_n} \zeta_{c,k,m,n} = 1- R_{c,k,m,n}, \IEEEyesnumber \label{eq:rateAWMMSE} \IEEEyessubnumber \label{eq:rateAWMMSE_1} \\ \zeta^{\text{MMSE}}_{k,q} &\triangleq& \min_{u^k_q,g^k_q} \zeta_{k,q} = 1- R_{k,q}. \IEEEyessubnumber \label{eq:rateAWMMSE_2}
    \vspace{-0.1cm}
\end{IEEEeqnarray}
\begin{figure}[t]
\begin{minipage}{1\linewidth}
\begin{algorithm}[H] 
 	\begin{algorithmic}[1] 
 	\STATE      
        \textbf{Initialize:} $n \leftarrow 0$, $\mathbf{P}^{[s]}$, $\mathbf{p}_c^{[s]}$, $\text{SR}^{[s]}$
 	\REPEAT 
        \STATE $n \leftarrow n+1$; 
        \STATE $\mathbf{P}^{[s-1]} \leftarrow \mathbf{P}^{[s]}$, \text{and} $\mathbf{p}_c^{[s-1]} \leftarrow \mathbf{p}_c^{[s]}$;
 	\STATE $\mathbf{U} \leftarrow \mathbf{U}^{\text{\ac{MMSE}}}(\mathbf{P}^{[s-1]},\mathbf{p}_c^{[s-1]})$;
 	\STATE $\mathbf{G} \leftarrow \mathbf{G}^{\text{\ac{MMSE}}}(\mathbf{P}^{[s-1]},\mathbf{p}_c^{[s-1]})$;
  \STATE update $(\bar{\boldsymbol{\zeta}^c},\mathbf{p}_c, \mathbf{P})$ by solving (\ref{eq:optProblem_2}) using the updated $\mathbf{U}$, and $\mathbf{G}$
 	\UNTIL convergence in $\text{SR}$ is reached 
    \end{algorithmic}
	\caption{Alternating Optimization for \ac{OFDM}-\ac{RSMA}} 
	\label{algo:OFDMRSMA} 
\end{algorithm}	
\end{minipage}
\vspace{-0.4cm}
\end{figure}
The following sets are defined to denote optimum \ac{MMSE} equalizers and weights: 
\vspace{-0.1cm}
\begin{IEEEeqnarray}{rCl}
   \mathbf{G}^{\text{\ac{MMSE}}} &\triangleq& \left\{ \left(g^{c,k,m}_n\right)^{\text{\ac{MMSE}}}, \left(g^k_q\right)^{\text{\ac{MMSE}}} \right \}, \nonumber \\
   \mathbf{U}^{\text{MMSE}} &\triangleq& \left\{ \left(u^{c,k,m}_n\right)^{\text{MMSE}},\left(u^k_q\right)^{\text{MMSE}}\right \}, \nonumber
   \vspace{-0.1cm}
\end{IEEEeqnarray}
where $k \in \mathcal{K}, \; m \in \mathcal{M}, \; n \in \mathcal{N}_c, \; q \in \mathcal{N}_p$. The deterministic version of AWMSE minimization problem can be formulated as follows: 
\vspace{-0.2cm}
\begin{mini}[3]<b>
{\boldsymbol{\zeta}^c,\mathbf{P}_c,\mathbf{P},\mathbf{U},\mathbf{G}}
{\sum^K_{k=1} \sum^M_{m=1} \sum_{n=1}^{N_c} \zeta_{c,k,m,n} + \sum^K_{k=1} \sum_{q=1}^{N_p} \zeta_{k,q}}
{}{}
\addConstraint{\sum_{m=1}^{M} \sum_{n=1}^{N_c} \zeta_{c,k,m,n}}{\leq \sum_{k=1}^K \sum_{m=1}^M  \sum_{n=1}^{N_c} X_{c,k,m,n} - MN_c(K-1),}
\addConstraint{}{\qquad \qquad \qquad \qquad \forall k \in \mathcal{K},}
\addConstraint{\Vert \mathbf{P} \Vert^2_F + \Vert \mathbf{P}_c \Vert_F^2}{\leq P_t, \labelOP{eq:optProblem_2}}
\addConstraint{\sum_{m=1}^{M} \sum_{n=1}^{N_c} \zeta_{c,k,m,n} +  \sum_{q=1}^{N_p} \zeta_{k,q}}{\leq M N_c + N_p - R_k^{\text{min}},}
\vspace{-0.1cm}
\end{mini}
where $X_{c,k,m,n} = 1-C_{c,k,m,n}$ and the transformation of the common rate can be written as follows: 
\begin{IEEEeqnarray}{rCl}
  \boldsymbol{\zeta}^c &=& \begin{bmatrix}
\zeta_{c,1,1,1} & \ldots & \zeta_{c,1,m,n} & \ldots & \zeta_{c,1,M,N_c} \nonumber \\
\vdots  & \vdots  & \ddots & \vdots    \nonumber \\
\zeta_{c,k,1,1} & \ldots & \zeta_{c,k,m,n} & \ldots & \zeta_{c,k,M,N_c} \nonumber \\
\vdots  & \vdots  & \ddots & \vdots    \nonumber \\
\zeta_{c,K,1,1} & \ldots & \zeta_{c,K,m,n} & \ldots & \zeta_{c,K,M,N_c} \nonumber 
\end{bmatrix}.
\end{IEEEeqnarray}
It can be seen that solving (\ref{eq:optProblem_2}) with respect to $\mathbf{U}$, and $\mathbf{G}$ leads to \ac{MMSE} solution of $\mathbf{G}^{\text{\ac{MMSE}}}$ $\mathbf{U}^{\text{\ac{MMSE}}}$ formed by corresponding \ac{MMSE} equalizers and weights. They satisfy the \ac{KKT} optimality conditions of (\ref{eq:optProblem_2}) for $\mathbf{P}$ and $\mathbf{P}_c$. For any point $(\boldsymbol{\zeta}^{c^\ast},\mathbf{P}_c^{\ast},\mathbf{P}^{\ast},\mathbf{U}^{\ast},\mathbf{G}^{\ast})$ satisfying the \ac{KKT} optimality conditions of (\ref{eq:optProblem_2}), the solution satisfies the \ac{KKT} optimality conditions of (\ref{eq:achievableRatesOFDM}). 
Then, the sum-rate problem in (\ref{eq:achievableRatesOFDM}) is transformed into \ac{WMMSE} problem of (\ref{eq:optProblem_2}). The optimization problem (\ref{eq:optProblem_2}) is still non-convex for the joint optimization with respect to parameters. It is shown that  when $(\boldsymbol{\zeta}^c, \mathbf{P}_c, \mathbf{P},\mathbf{U})$ are fixed, the optimal equalizer is the \ac{MMSE} equalizer $\mathbf{G}^{\text{MMSE}}$, and when $(\boldsymbol{\zeta}^c, \mathbf{P}_c, \mathbf{P},\mathbf{G})$ are fixed, the optimal weight is the \ac{MMSE} weight $\mathbf{U}^{\text{MMSE}}$. When $(\mathbf{U},\mathbf{G})$ are fixed, $(\boldsymbol{\zeta}^c,\mathbf{P}_c, \mathbf{P})$ is coupled in the optimization
problem (\ref{eq:optProblem_2}), closed-form solution cannot be derived. However, it is a convex \ac{QCQP} which can be solved using interior-point methods. These properties lead us to use alternating optimization to solve the problem. In $s$th iteration of the alternating optimization algorithm, the equalizers and weights are firstly updated using the power allocation matrix obtained in the ($s-1$)th iteration $(\mathbf{U}, \mathbf{G}) =
(\mathbf{U}^{\text{\ac{MMSE}}}(\mathbf{P}^{[s-1]},\mathbf{P}_c^{[s-1]}), \mathbf{G}^{\text{\ac{MMSE}}}(\mathbf{P}^{[s-1]},\mathbf{P}_c^{[s-1]}))$. $(\boldsymbol{\zeta}^c, \mathbf{P}_c, \mathbf{P})$ can  be updated by solving the problem (\ref{eq:optProblem_2}), with the updated $(\mathbf{U}, \mathbf{G})$.  $(\boldsymbol{\zeta}^c, \mathbf{P}_c, \mathbf{P})$ and $(\mathbf{U}, \mathbf{G})$ are iteratively updated until the sum-rate converges. The details of the alternating optimization algorithm is shown in Algorithm \ref{algo:OFDMRSMA}, where $\text{SR}^{[s]}$ is the SR calculated based on the updated $(\boldsymbol{\zeta}^c, \mathbf{P}_c, \mathbf{P})$ in $s$th iteration. The alternating optimization algorithm is guaranteed to converge as sum-rate increasing in each iteration and it is bounded above for a given power constraint. Two steps are followed in every iteration of alternating optimization which are explained below. 

\subsubsection{Fix the precoder and update equalizers and weights}
The equalizers and weights are updated as $(\mathbf{G},\mathbf{U}) = (\mathbf{G}^{\text{MMSE}}(\mathbf{P}^{[s-1]}_c,\mathbf{P}^{[s-1]})$, $\mathbf{U}^{\text{MMSE}}(\mathbf{P}^{[s-1]}_c,\mathbf{P}^{[s-1]}))$ at the $s$th iteration of the algorithm, where $\mathbf{P}^{[s-1]}_c,\mathbf{P}^{[s-1]}$ are the common and private precoders found at the $(s-1)$th iteration. The intermediate parameters which are obtained using the updated $(\mathbf{G}, \mathbf{U})$ can be listed as follows: 
\vspace{-0.2cm}
\begin{IEEEeqnarray}{rCl}
   \boldsymbol{\alpha}_{c,k,m} & = & \mathbf{u}^{c,k,m} \odot \left\vert \mathbf{g}^{c,k,m}  \right \vert^{o 2}, \nonumber
   \\ \boldsymbol{\alpha}_{k} & = & \mathbf{u}^k \odot \left \vert \mathbf{g}^k \right \vert^{o 2}, \nonumber \\ \boldsymbol{\beta}_{c,k,m} &=&  \diag{\left(\boldsymbol{\alpha}_{c,k,m}\right)^{o (1/2)}}  \mathbf{F}_c \mathbf{B}_{c,m} \mathbf{H}_k \mathbf{A}_{c,m} \mathbf{F}_c^H, \nonumber
    \\ \boldsymbol{\beta}_{c,p,k,m} &=&  \diag{\left(\boldsymbol{\alpha}_{c,k,m}\right)^{o (1/2)}}   \mathbf{F}_c \mathbf{B}_{c,m} \mathbf{H}_k \mathbf{A}_p \mathbf{F}_p^H,  \nonumber
    \\
    \boldsymbol{\beta}_{k}  &=& \diag{\left(\boldsymbol{\alpha}_{k}\right)^{o (1/2)}}  \mathbf{F}_p \mathbf{B}_p \mathbf{H}_k \mathbf{A}_p \mathbf{F}_p^H, \nonumber
   \\ \boldsymbol{f}_{c,k,m} &=& \diag{\mathbf{u}^{c,k,m} \odot \mathbf{g}^{c,k,m}}  \mathbf{F}_c \mathbf{B}_{c,m} \mathbf{H}_k \mathbf{A}_{c,m} \mathbf{F}_c^H, \nonumber
    \IEEEeqnarraynumspace \\
    \boldsymbol{f}_{k} &=& \diag{\mathbf{u}^{k} \odot \mathbf{g}^k}  \mathbf{F}_p \mathbf{B}_p \mathbf{H}_k \mathbf{A}_p \mathbf{F}_p^H, \nonumber \IEEEeqnarraynumspace \\
    \boldsymbol{\upsilon}_{c,k,m}  & = & \log_2\left(\mathbf{u}^{c,k,m}\right), \quad \text{and} \quad \boldsymbol{\upsilon}_{k} = \log_2\left(\mathbf{u}^k \right). \nonumber
\end{IEEEeqnarray}

\subsubsection{Fix the equalizer and weights update the precoder}

The optimization problem can be written as follows:
\begin{mini}[3]<b>
{\substack{\boldsymbol{\zeta}^c,\\\mathbf{P}_c,\\\mathbf{P}}}{\sum_{k=1}^K \Bigg[\sum_{m=1}^{M} \sum_{n=1}^{N_c} \zeta_{c,k,m,n} + \sum_{q=1}^{N_p} \Bigg(\sum_{\substack{i=1 \\ i \neq k}}^K \sum_{j=1}^{N_p} \vert\chi^i_{q,j}\vert^2  + \vert \kappa^{k,q}_{q,q} \vert^2}
{}{}
\breakObjective{ + \sum_{j=1}^{N_p} \vert \bar{\kappa}^{k,q}_{q,j}\vert^2 + \left(\boldsymbol{\alpha}_{k}\right)_q \sigma^2  - 2\mathfrak{R}\left\{ \omega^k_q \right\} + u^k_q - \upsilon^k_q \Bigg) \Bigg]}
\addConstraint{ \sum_{m=1}^M \sum_{n=1}^{N_c} \Bigg( \sum_{i=1}^K  \sum_{q=1}^{N_p} \vert \psi^{i,m}_{n,q} \vert^2 +  \vert \phi^{k,m,n}_{n,n} \vert^2 + \sum_{j=1}^{N_c} \vert \bar{\phi}^{k,m,n}_{n,j} \vert^2  }{}
\addConstraint{+\left(\boldsymbol{\alpha}_{c,k,m}\right)_n + \sigma^2  - 2\mathfrak{R}\left\{ \omega^{c,k}_{n}\right\} }{}
\addConstraint{  + u^{c,k}_n + v^{c,k}_n \Bigg)}{\leq \sum_{k=1}^K \sum_{m=1}^M \sum_{n=1}^{N_c} \zeta_{c,k,m,n},}
\addConstraint{\Vert \mathbf{P} \Vert^2_F + \Vert \mathbf{p}_c \Vert^2}{\leq P_t, \labelOP{eq:UpdateWeights}}
\addConstraint{\sum_{m=1}^{M} \sum_{n=1}^{N_c} \zeta_{c,k,m,n} +  \sum_{q=1}^{N_p} \zeta_{k,q}}{\leq M N_c + N_p - R_k^{\text{min}}},
\end{mini}
where
\vspace{-0.2cm}
\begin{IEEEeqnarray}{rCl} 	
    \boldsymbol{\chi}^i &=& \boldsymbol{\beta}_i \diag{\mathbf{p}_i}, \quad \forall i \in \mathcal{K},  \nonumber \\ 
    \boldsymbol{\kappa}^{k,q} &=& \boldsymbol{\beta}_k \diag{\mathbf{p}_{k,q}}, \quad \forall k \in \mathcal{K} \; , \forall q \in \mathcal{N}_p, \nonumber \\ \bar{\boldsymbol{\kappa}}^{k,q} &=& \boldsymbol{\beta}_k \diag{\bar{\mathbf{p}}_{k,q}}, \quad \forall k \in \mathcal{K} \; , \forall q \in \mathcal{N}_p, \nonumber \\ \boldsymbol{\omega}^k &=& \diag{\mathbf{f}_{k}\diag{ \mathbf{p}_k}}, \quad \forall k \in \mathcal{K}, \nonumber \\ 
    \boldsymbol{\psi}^{i,m} &=& \boldsymbol{\beta}_{c,p,i,m} \diag{\mathbf{p}_i}, \quad \forall i \in \mathcal{K},  \nonumber \\ 
    \boldsymbol{\phi}^{k,m,n} &=& \boldsymbol{\beta}_{c,k,m} \diag{\mathbf{p}_{c,n}}, \quad \forall k \in \mathcal{K} \; , \forall n \in \mathcal{N}_c, \nonumber \\ \bar{\boldsymbol{\phi}}^{k,m,n} &=& \boldsymbol{\beta}_{c,k,m} \diag{\bar{\mathbf{p}}_{c,n}}, \quad \forall k \in \mathcal{K} \; , \forall n \in \mathcal{N}_c,  \nonumber \\ \boldsymbol{\omega}^{c,k,m} &=&\diag{\mathbf{f}_{c,k,m} \diag{\mathbf{p}_{c,m}}}, \quad \forall k \in \mathcal{K}. \nonumber
\end{IEEEeqnarray}

In the Section \ref{sec:Results}, numerical evaluations for the proposed \ac{OFDM}-\ac{RSMA} scheme are demonstrated where convex optimization problems are solved using the CVX toolbox \cite{cvx}. 

\section{\ac{OFDM}-\ac{NOMA} Transmission Scheme}\label{sec:OFDM-NOMA}

In the $K$-user \ac{OFDM}-\ac{NOMA} scheme, \ac{SIC} order for users is not changed throughout one \ac{OFDM} symbol. Subcarrier-based \ac{SIC} is not practical since coded-\ac{OFDM} requires the \ac{LLR} values across all subcarriers before decoding the user's intended message. The $k$th user message, $\mathbf{x}_k$ is decoded after $k-1$ users' signal are successively canceled. Unlike the \ac{OFDM}-\ac{RSMA} scheme explained in Sec. \ref{sec:OFDMRSMAtransmission}, $\mathbf{x}_k$ is designed to convey information of $k$th user only and there is only one stream dedicated to $k$th user after removing $k-1$ users' signal. Let the set $\mathcal{M}_k = \{1, \ldots, M_k \}$ define the number of \ac{OFDM} symbol transmitted by $k$th user in a fixed $T$ duration where $\mathcal{M}_k \in \mathcal{M} = \{ \mathcal{M}_1 , \ldots, \mathcal{M}_K \}$.

At the $k$th user's receiver, the superimposed signal $\mathbf{y}_k$ is processed with \ac{OFDM} demodulator after removing all $k-1$ user's signal whose decoding order is earlier than $k$th user. The received frequency domain signal at the $k$th user's $m$th \ac{OFDM} symbol, $\mathbf{r}_{k,m}$, can be written as follows:
\vspace{-0.1cm}
\begin{IEEEeqnarray}{rCl} 
    \mathbf{r}_{k,m} &=& \mathbf{F}_k \mathbf{B}_{k,m} \left(\mathbf{y}_{k} - \mathbf{H}_k \sum_{l=1}^{k-1} \sum_{\eta=1}^{M_l} \mathbf{A}_{l,\eta} \mathbf{F}_l^H \diag{\mathbf{p}_{l,\eta}}\mathbf{d}_{l,\eta} \right) \nonumber	\\ 
    \label{eq:nomaReceived}
    \vspace{-0.1cm}
\end{IEEEeqnarray}
where subscripts $k$ and $m$ are used in order to note that users may utilize different \ac{SCS} value. For the certain channel state, the average received power, $T_{k,m,q} \triangleq \mathsf{E} \left\{ \left\vert {(\mathbf{r}_{k})}_q \right \vert^2 \right\}$, at the $q$th subcarrier of $\mathbf{r}_{k,m}$ can be written as follows:
\vspace{-0.15cm}
\begin{IEEEeqnarray}{rCl} 
    T_{k,m,q} & = & \left \vert v^{k,m,q}_{q,q} \right \vert^2 + \underbrace{ \sum_{j=1}^{N_k} \left \vert \bar{v}^{k,m,q}_{q,j} \right \vert^2 + \sum_{\substack{i=k+1}}^{K} \sum_{j=1}^{N_k} \left \vert w^{k,i,m}_{q,j}  \right \vert^2 + \sigma^2}_{I_{k,m,q}}, \nonumber \\ \label{eq:NOMApower} 
    \vspace{-0.2cm}
\end{IEEEeqnarray}	
where $N_k$ is the total subcarrier number of $k$th user and
\vspace{-0.15cm}
\begin{IEEEeqnarray}{rCl} 	
    \mathbf{V}^{k,m,q} &=& \mathbf{F}_k \mathbf{B}_{k,m} \mathbf{H}_k \mathbf{A}_{k,m} \mathbf{F}_k^H \diag{p_{k,m,q}\mathbf{e}_q}, \nonumber  \IEEEeqnarraynumspace \\
    \bar{\mathbf{V}}^{k,m,q} &=& \mathbf{F}_k \mathbf{B}_{k,m} \mathbf{H}_k \mathbf{A}_{k,m} \mathbf{F}_k^H \diag{\bar{\mathbf{p}}_{k,m,q}},  \nonumber \IEEEeqnarraynumspace \\
    \mathbf{W}^{k,i,m} &=& \mathbf{F}_k \mathbf{B}_{k,m} \mathbf{H}_k \mathbf{A}_i \mathbf{F}_i^H \diag{\mathbf{p}_i}, \; \forall i \in \{k+1, \ldots, K\}, \nonumber \IEEEeqnarraynumspace
    \vspace{-0.15cm}
\end{IEEEeqnarray}
for the subcarrier set of $\mathcal{N}_k = \{1,2,\ldots,N_k \}$. The vector $\bar{\mathbf{p}}_{k,m,q}$ denotes that $q$th subcarrier of $m$th \ac{OFDM} symbol is forced not to carry any energy, i.e.,  $\bar{\mathbf{p}}_{k,q} = [p_{k,1},\ldots,p_{k,{q-1}},0,p_{k,{q+1}},\ldots,p_{k,{N_k}}]^T$.


The optimization problem for achievable rate maximization using \ac{OFDM}-\ac{NOMA} can be formulated as follows:
\vspace{-0.15cm}
\begin{maxi}
{\mathbf{P}}{ \sum^K_{k=1} \sum^{M_k}_{m=1} \sum_{q=1}^{N_k} R_{k,m,q}}
{\label{eq:NOMAoptProblem}}{}
\addConstraint{\Vert \mathbf{P} \Vert^2_F}{\leq P_t},
\addConstraint{R_k}{\geq R_k^{\text{min}}},
\vspace{-0.3cm}
\end{maxi}
where the matrix $\mathbf{P} = [\mathbf{p}_{1,1},\ldots,\mathbf{p}_{1,M_1},\ldots,\mathbf{p}_{1,M_K},\ldots,\mathbf{p}_{K,M_K}]$ is defined as the collection of all users' precoding vectors, $P_t$ is the total transmit power, and $R_k^{\text{min}}$ is the minimum data rate constraint assigned to $k$th user. The estimate of $m$th \ac{OFDM} symbol of $k$th user signal in the frequency domain, $\hat{\mathbf{d}}_{k,m}$, can be obtained 
employing one-tap equalizer vector, $\mathbf{g}^{k,m}$, on $\mathbf{r}_{k,m}$ given in (\ref{eq:nomaReceived}). Similar to proposed \ac{OFDM}-\ac{RSMA} scheme, rate-AWMMSE transformation can be applied to solve \ac{OFDM}-\ac{NOMA} sum-rate optimization problem given in (\ref{eq:NOMAoptProblem}).

\vspace{-0.2cm}
\section{Simulation Results}
\label{sec:Results}
In this section, the performance of \ac{OFDMA}, \ac{OFDM}-\ac{NOMA}, and \ac{OFDM}-\ac{RSMA} is evaluated in a wide variety of channel deployments, such as different channel strengths, Delay and Doppler spread. Several analysis such as achievable rate, fairness, power allocation over different channel conditions in the two-user case are illustrated. Throughout the numerical evaluations, it is assumed that the users and \ac{BS} are able to estimate the channel perfectly when they try to design subcarrier and power allocation at the \ac{BS} and equalizers at the users side. Table \ref{tab:simulationParameters} shows the simulation parameters that are used throughout the study. We consider a scenario with $K=2$, where generalization to more users case can be easily done. Two different \acp{SCS} in \ac{CP}-\ac{OFDM} waveform is used which are \SI{140}{\kilo\hertz} or \SI{60}{\kilo\hertz}. Correspondingly, 15 or 35 subcarriers is used in an \ac{OFDM} symbol whose bandwidth is fixed to \SI{2.1}{\mega\hertz}. 
\begin{table}[h]
\vspace{-0.3cm}
\small
\caption{Simulation Parameters}
\label{tab:simulationParameters}
\centering
\vspace{-0.2cm}
\begin{tabular}{c|c}
\hline
\bfseries Parameter & \bfseries Value\\
	\hline\hline
		User Number $K$ & 2 \\
		\hline
		\Ac{SCS} & \SI{60}{\kilo\hertz} or \SI{140}{\kilo\hertz} \\
		\hline
		Subcarrier number  & 35 or 15 \\
		\hline
		The frame duration & \SI{19.04}{\micro\second} \\
		\hline
		\ac{CP} length in sample  & 5 \\ 
  \hline 
  CP length duration & \SI{2.4}{\micro\second}
  \\
  \hline           Sampling rate \& Total BW & \SI{2.1}{\mega\hertz}
		 \\ \hline
  Maximum Delay spread & \SI{2}{\micro\second} \\ \hline Maximum Doppler spread & \SI{12}{\kilo\hertz} \\ \hline
\end{tabular}
\vspace{-0.4cm}
\end{table}

Since the sum-rate optimization problems formed before are non-convex, the initialization of subcarrier and power allocation values, $\mathbf{P}$ and $\mathbf{P}_c$, is crucial to find the reliable result. For \ac{OFDMA}, subcarriers are allocated to users with respect to channel power on corresponding subcarrier. The allocation is done in subcarrier based and orthogonal manner. Besides, power allocation is done with water-filling algorithm after assigning subcarrier to the corresponding user \cite{Tse_fundamentalsOfWirelessComm_2005}. The similar water-filling procedure is followed to initialize the power and subcarrier allocation for \ac{OFDM}-\ac{NOMA} and \ac{OFDM}-\ac{RSMA} techniques. It is observed that water-filling based power initialization gives the highest sum-rate among other initialization techniques under the frequency selective channel. 

\begin{figure}[t]
\vspace{-0.2cm}
    \centering
    \subfloat[]{\includegraphics[width=0.8\linewidth]{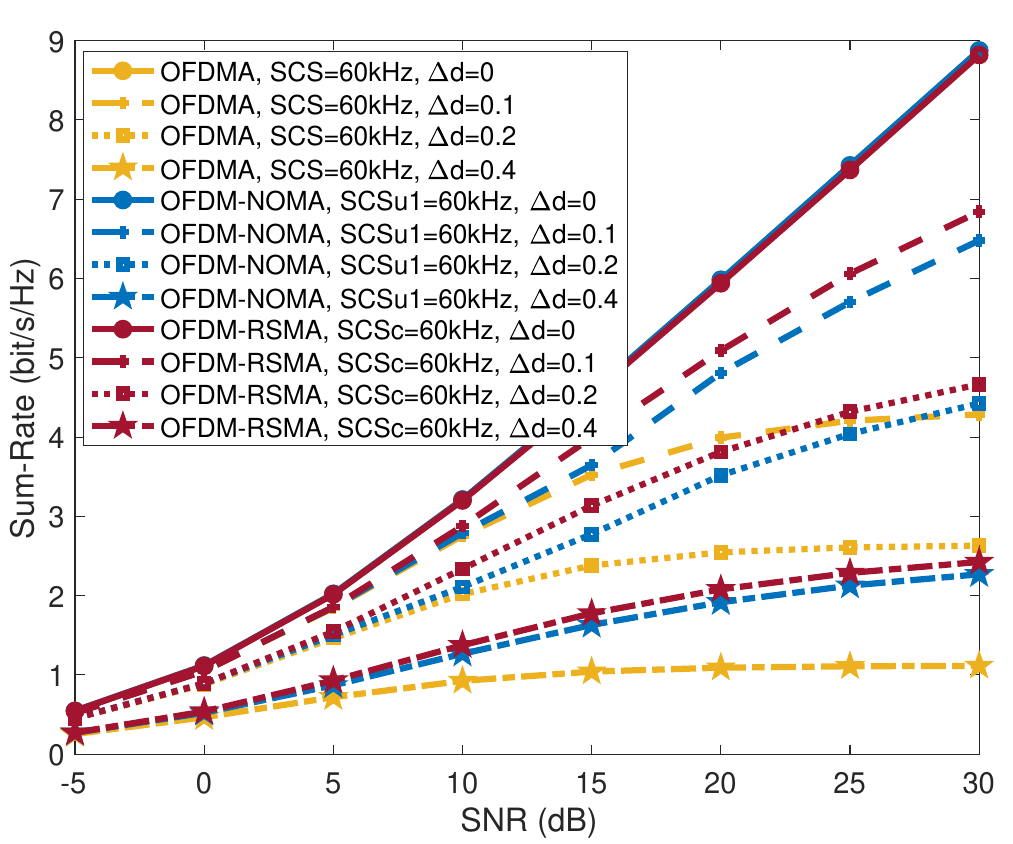} \label{fig:dataRateDoublySelective}} \vspace{-0.3cm} \hfill
    \subfloat[]{\includegraphics[width=0.8\linewidth]{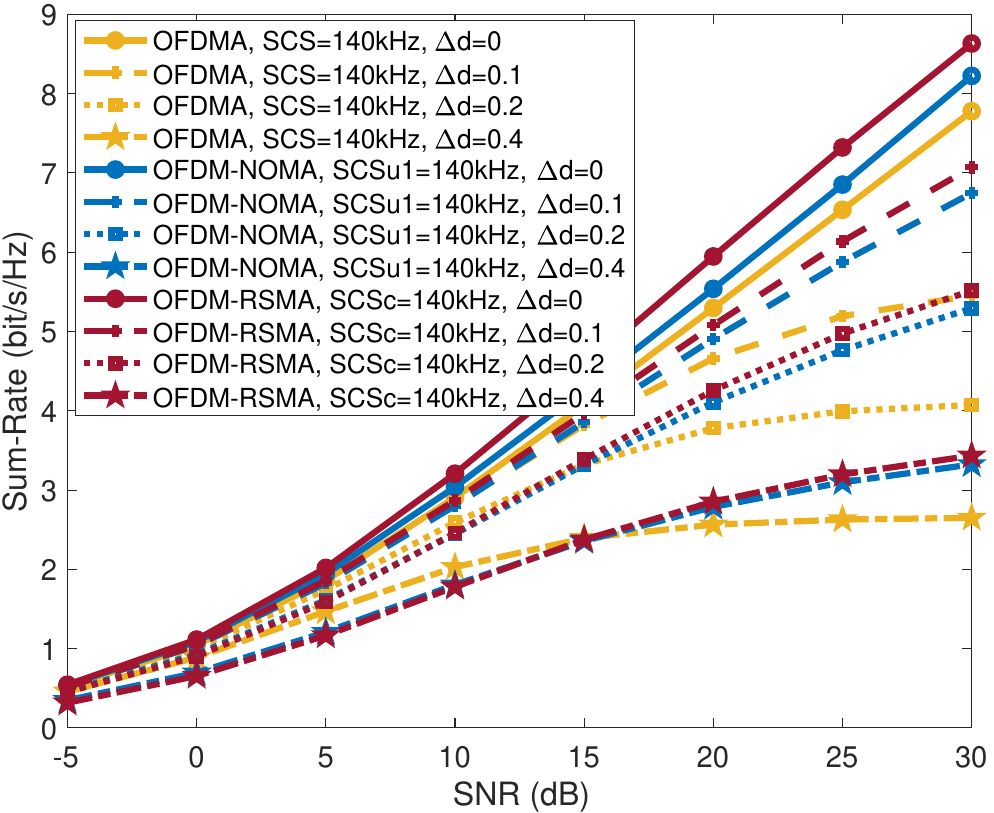} \label{fig:dataRateDoublySelectiveLargeSCS}} \vspace{-0.3cm} \hfill
    \subfloat[]{\includegraphics[width=0.8\linewidth]{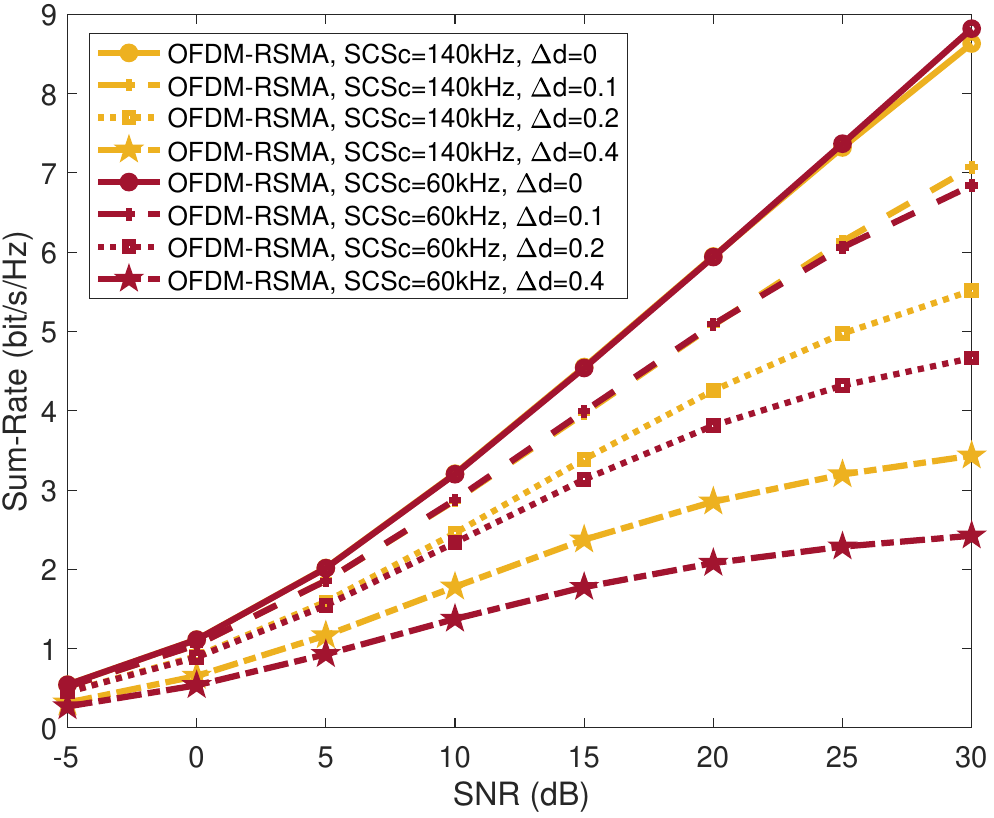} \label{fig:dataRateDiffRSMAcommonSCS}} \vspace{-0.2cm}
    \caption{Sum-rate analysis of proposed \ac{OFDM}-\ac{RSMA} scheme compared to conventional \ac{OFDMA} and \ac{OFDM}-\ac{NOMA} under the frequency selective and doubly selective channel (a) \acp{SCS} of all streams are set to \SI{60}{\kilo\hertz}, (b) \acp{SCS} of both users in \ac{OFDMA}, first user in \ac{OFDM}-\ac{NOMA}, and common stream in \ac{OFDM}-\ac{RSMA} are set to \SI{140}{\kilo\hertz} where user-2 in \ac{OFDM}-\ac{NOMA} and private streams in \ac{OFDM}-\ac{RSMA} have \SI{60}{\kilo\hertz} \ac{SCS}, (c) different \acp{SCS} for common stream in OFDM-RSMA when \ac{SCS} of private streams is \SI{60}{\kilo\hertz}.}
    \label{fig:dataRateComparison} 
    \vspace{-0.4cm}
\end{figure}
\begin{figure*}[t]
    \centering
    \subfloat[]{
\includegraphics[width=0.8\linewidth]{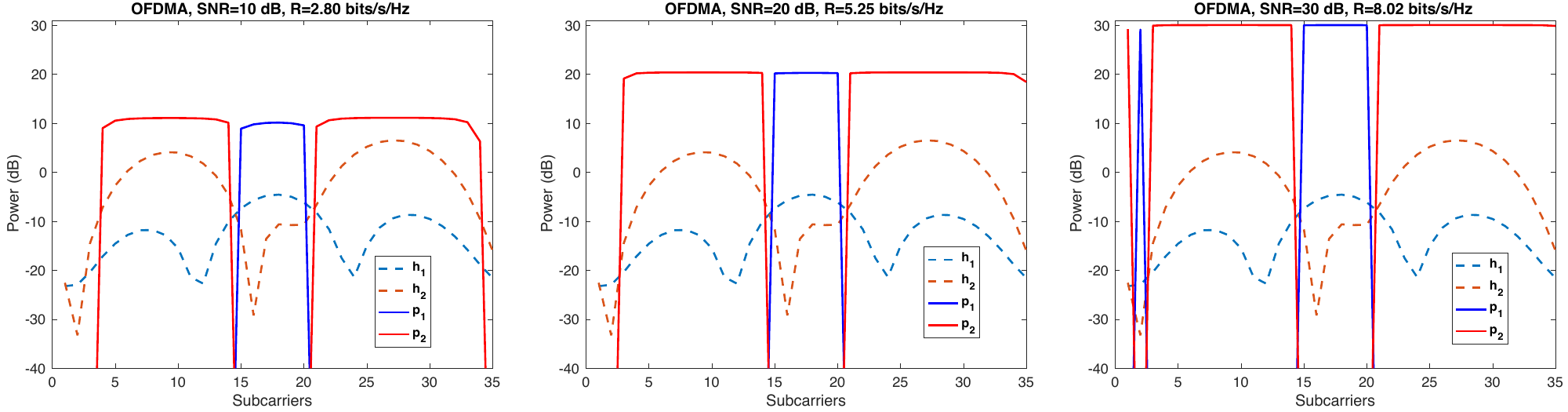}
\label{fig:powerBySubcarriersOFDMA}} \vspace{-0.2cm}\hfill
\subfloat[]{
\includegraphics[width=0.8\linewidth]{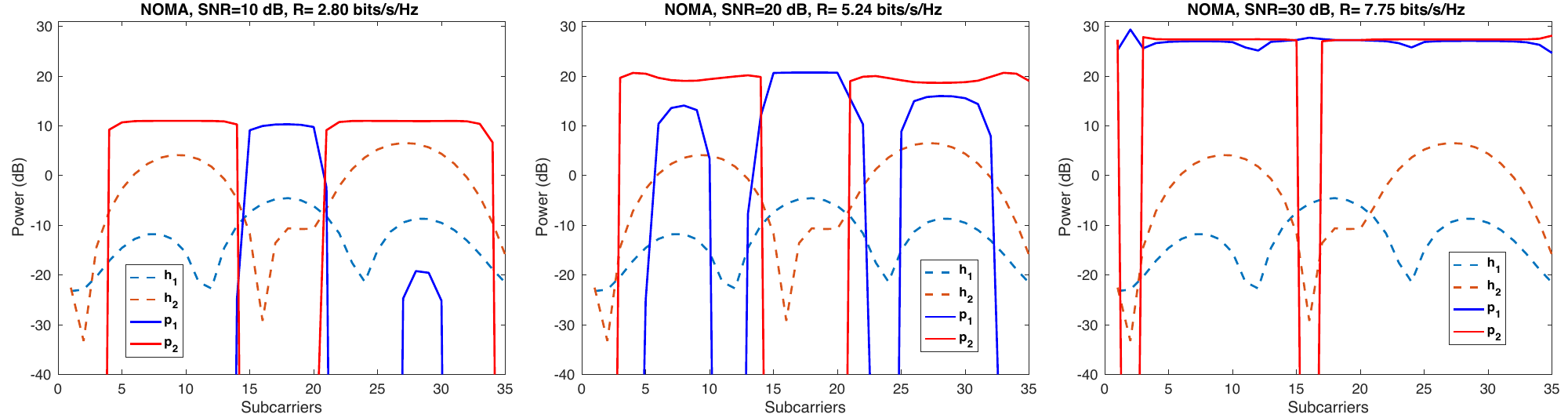}
\label{fig:powerBySubcarriersNOMA}
} \vspace{-0.2cm} \hfill
\subfloat[]{
\includegraphics[width=0.8\linewidth]{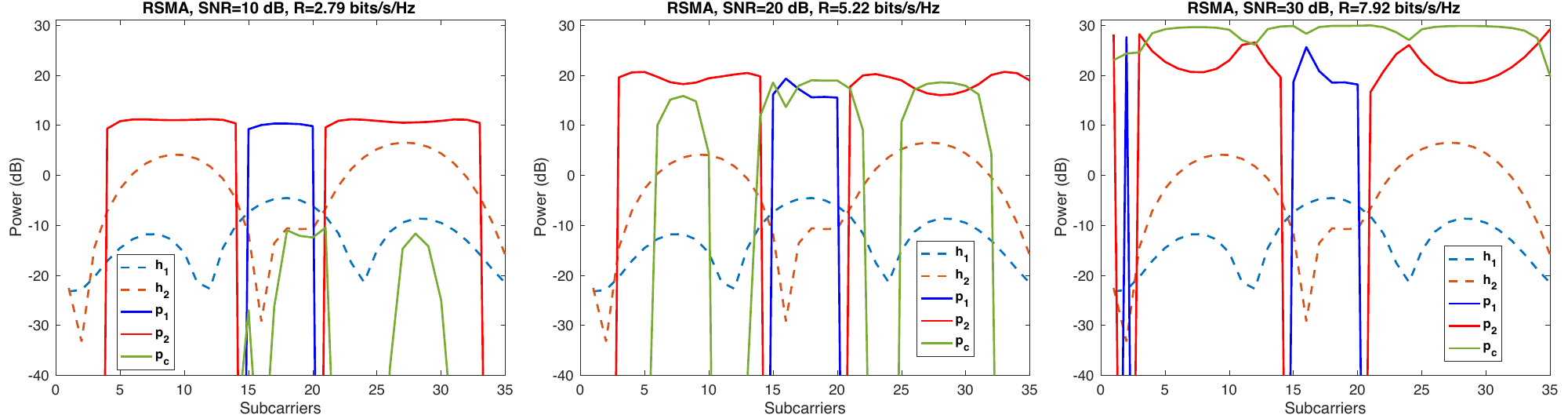}
\label{fig:powerBySubcarriersRSMA}
} \vspace{-0.2cm}  \caption{Power allocation across the \ac{OFDM} subcarriers under various \ac{SNR} levels in frequency selective channel, (a) conventional OFDMA, (b) \ac{OFDM}-\ac{NOMA}, (c) \ac{OFDM}-\ac{RSMA}.} 
\label{fig:powerBySubcarriers}
\vspace{-0.5cm}
\end{figure*}

Fig. \ref{fig:rsmaGeneralization} shows the OFDM based \ac{MA} schemes studied throughout the paper including the conventional \ac{OFDMA}, \ac{OFDM}-\ac{NOMA}, and the proposed \ac{OFDM}-\ac{RSMA}. Each scheme serves two users. Multi-numerology \ac{OFDM} concept is also investigated to make proposed \ac{MA} schemes more robust in the presence of time selectivity. Fig. \ref{fig:OFDMAschemes1} and Fig. \ref{fig:OFDMAschemes2} demonstrate two \ac{OFDMA} schemes with \SI{60}{\kilo\hertz} \ac{SCS} and \SI{140}{\kilo\hertz} \ac{SCS}, respectively. During an \ac{OFDM} frame duration $T$, latter scheme carries two \ac{OFDM} symbols whereas the former has only one \ac{OFDM} symbol. Two different \ac{OFDM}-\ac{NOMA} configurations are depicted in Fig. \ref{fig:NOMAschemes1} and \ref{fig:NOMAschemes2}.  The \ac{SCS} of users in the first configuration is same and set as \SI{60}{\kilo\hertz}. In the scheme shown in Fig. \ref{fig:NOMAschemes2}, the \ac{SCS} of user-1 is widened to \SI{140}{\kilo\hertz} and user-2's \ac{SCS} remains the same as \SI{60}{\kilo\hertz}. It is assumed that the user-1 is the weak user having a smaller overall channel gain than the user-2 (strong user), therefore, user-1 signal is decoded first in the \ac{SIC} process. Finally, the proposed \ac{OFDM}-\ac{RSMA} structure is analyzed with two different configurations that can be seen in Fig. \ref{fig:RSMAschemes1} and \ref{fig:RSMAschemes2}. Among two configurations, only \ac{SCS} of common stream changes from \SI{60}{\kilo\hertz} to \SI{140}{\kilo\hertz} to make the scheme more robust against \ac{ICI}. The SCS of private streams are determined as \SI{60}{\kilo\hertz}. 

\subsubsection{Achievable Sum-Rate Analysis}
In Fig. \ref{fig:dataRateComparison}, the performance of \ac{OFDMA}, \ac{OFDM}-\ac{NOMA}, and \ac{OFDM}-\ac{RSMA} are investigated under the frequency selective channel with different Doppler spreads. Here, $\Delta d = \frac{f_d}{\Delta f}$ parameter corresponds to the maximum Doppler spread $(f_d)$ over the subcarrier spacing of \SI{60}{\kilo\hertz} $(\Delta f = \SI{60}{\kilo\hertz})$. Four cases ($\Delta d = 0$, $\Delta d = 0.1 $, $\Delta d = 0.2$, and $\Delta d = 0.4$) are demonstrated in all plots together where $\Delta d = 0$ stands for the frequency selective channel model without any time selectivity.  

Fig. \ref{fig:dataRateDoublySelective}
demonstrates the performance of three schemes where all streams have \SI{60}{\kilo\hertz} \ac{SCS} in \ac{OFDM} waveform. For \ac{OFDMA}, the \ac{ICI} due to Doppler spread makes the sum-rate curve to saturate for high \ac{SNR} regime which can be regarded as interference dominated regime. On the other hand, \ac{OFDM}-\ac{NOMA} has robustness to \ac{ICI} because the interference is also decoded and removed when SIC operation is done to remove user-1 signal. Rather, \ac{OFDM}-\ac{RSMA} clearly outperforms both \ac{OFDMA} and \ac{OFDM}-\ac{NOMA} in terms of sum-rate gain over all Doppler spread regimes. Utilizing common stream, which introduces data rate to both users, allows us to exploit all power variations of frequency fading channel in \ac{OFDM} waveform. Also, it provides robustness to ICI mitigation since the interference is partially decoded and partially treated as noise in \ac{OFDM}-\ac{RSMA} scheme. With RS, the amount of interference decoded by both users (through the presence
of common stream) is adjusted dynamically to the
channel conditions (channel strength on subcarriers and leakage due to Doppler spread).

Fig. \ref{fig:dataRateDoublySelectiveLargeSCS}
demonstrates the performance of three schemes when both users in \ac{OFDMA}, user-1 in \ac{OFDM}-\ac{NOMA} and common stream in \ac{OFDM}-\ac{RSMA} have larger \ac{SCS} ($\Delta f = \SI{140}{\kilo\hertz}$); whereas user-2 in \ac{OFDM}-\ac{NOMA} and private streams in \ac{OFDM}-\ac{RSMA} have narrower \ac{SCS} ($\Delta f = \SI{60}{\kilo\hertz}$). Since the \ac{SCS} of \ac{OFDMA} is wider, it can reach higher data rate compared to Fig. \ref{fig:dataRateDoublySelective} under the presence of \ac{ICI}. However,  the sum-rate of \ac{OFDMA} keeps saturating in high SNR regime. This is because \ac{OFDMA} does not exploit the interference structure and treats interference as noise. In every regime, \ac{OFDM}-\ac{RSMA} outperforms \ac{OFDMA} and \ac{OFDM}-\ac{NOMA} owing to its robust interference management and efficient resource allocation. It can also be seen that additional overhead due to \ac{CP} lowers sum-rate for \ac{OFDMA} and \ac{OFDM}-\ac{NOMA} when there is no \ac{ICI} ($\Delta d = 0$).  

Fig. \ref{fig:dataRateDiffRSMAcommonSCS} shows the effect of different \ac{SCS} selection in the common stream of \ac{OFDM}-\ac{RSMA} on the achievable sum-rate performance. In the no Doppler case with high SNR regime where more power is allocated to common stream, three is a slight decrease in the sum-rate of the scheme with wider \ac{SCS} in the common stream. It is because of the fact that there are less data subcarriers in the common stream to carry information when its \ac{SCS} is wider. However, as \ac{ICI} power increases, the scheme with wider \ac{SCS} in the common stream outperforms the one with narrower \ac{SCS}. 

\subsubsection{Power Allocation Analysis}

Fig. \ref{fig:powerBySubcarriers} illustrates power allocation of studied \ac{MA} schemes over the \ac{OFDM} subcarriers in the presence of frequency-selective channel without any Doppler spread. Dashed lines show the channel gain of users whereas solid lines are the power levels allocated to corresponding streams. Conventional \ac{OFDMA} seen in Fig. \ref{fig:powerBySubcarriersOFDMA} applies well known waterfilling technique to maximize the sum-rate. The \ac{OFDM}-\ac{NOMA} scheme shown in Fig. \ref{fig:powerBySubcarriersNOMA} may prefer orthogonal transmission when one user experiences fading at corresponding subcarrier. Finally, power allocation over subcarriers for the proposed \ac{OFDM}-\ac{RSMA} is depicted in Fig.  \ref{fig:powerBySubcarriersRSMA} where the power of common stream increases as \ac{SNR} increases. Also, it can be seen that private streams follow the orthogonal transmission showing that \ac{OFDM}-\ac{RSMA} bridges between \ac{OFDMA} and \ac{OFDM}-\ac{NOMA}. 

\begin{figure}[t]
    \vspace{-0.2cm}
    \centering
    \subfloat[]{\includegraphics[width=0.8\linewidth]{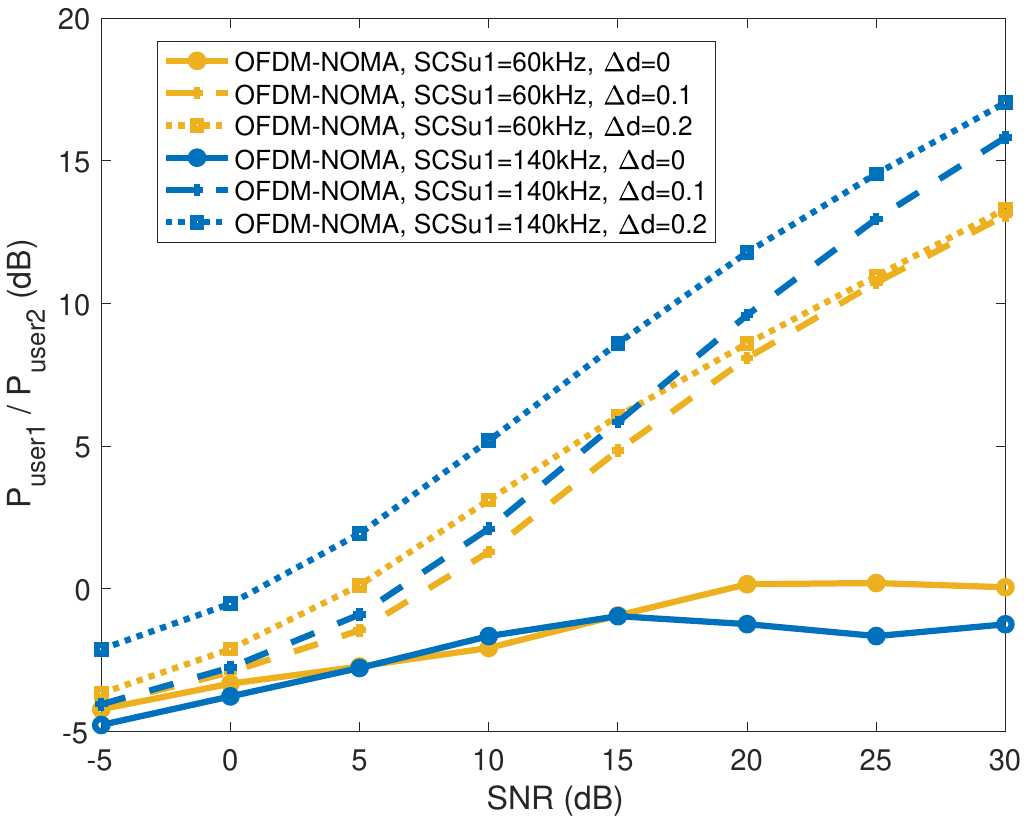} \label{fig:nomaPowerAnalysis}} \vspace{-0.2cm} \hfill
    \subfloat[]{\includegraphics[width=0.8\linewidth]{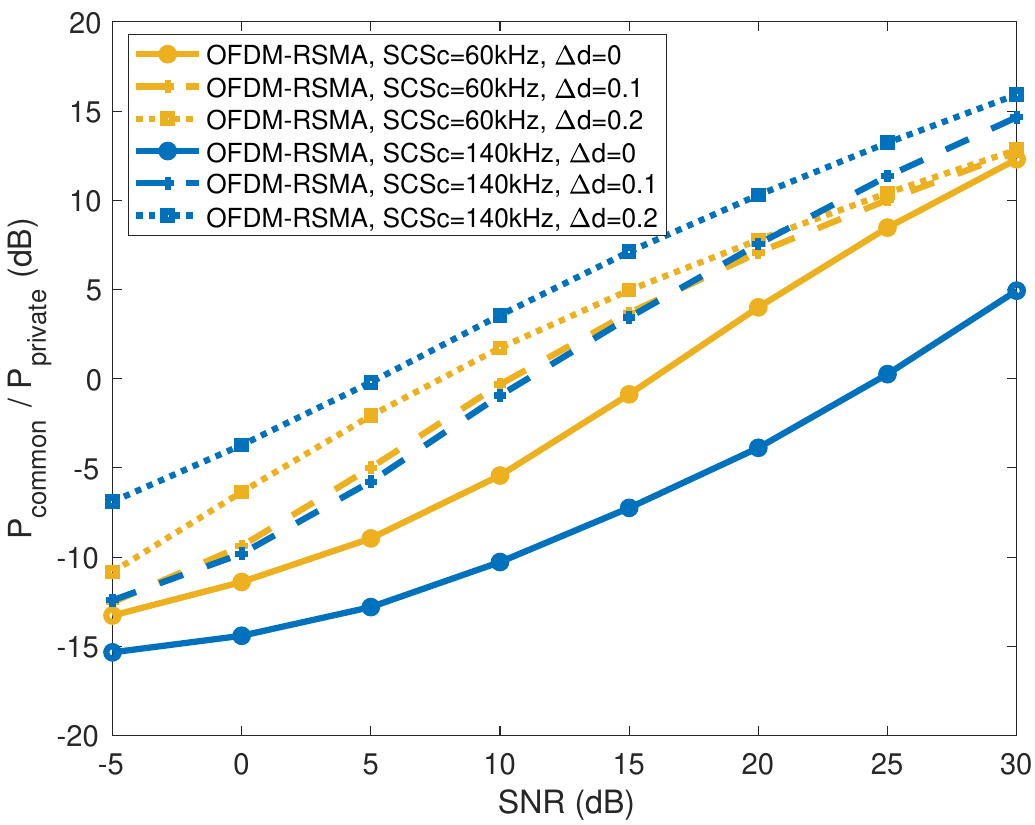}
    \label{fig:rsmaPowerAnalysis}} \vspace{-0.2cm}
    \caption{Power allocation analysis in \ac{OFDM}-\ac{NOMA} and \ac{OFDM}-\ac{RSMA} schemes, (a) power ratio between user 1 and user 2 in \ac{OFDM}-\ac{NOMA}, (b) power ratio between common stream and private streams in \ac{OFDM}-\ac{RSMA}.}
    \label{fig:powerAllocation} \vspace{-0.3cm}
\end{figure}

Fig. \ref{fig:powerAllocation} demonstrates the power allocation analysis in \ac{OFDM}-\ac{NOMA} and \ac{OFDM}-\ac{RSMA} after the sum-rate maximization problem is solved. Power difference between user-1 and user-2 for \ac{OFDM}-\ac{NOMA} can be seen in Fig. \ref{fig:nomaPowerAnalysis} whereas  power ratio between common and private streams for \ac{OFDM}-\ac{RSMA} is shown in Fig. \ref{fig:rsmaPowerAnalysis}. 
To gain interference robustness and prevent data loss, \ac{OFDM}-\ac{NOMA} allocates more power on user-1 in the interference dominated regime due to \ac{ICI}. As \ac{ICI} increases, the more power is allocated to the user-1 stream, as expected. Since the \ac{OFDM} stream with higher \ac{SCS} is exposed to less interference under doubly dispersive channel, more power is allocated to user-1 when its \ac{SCS} is at \SI{140}{\kilo\hertz}.  On the other hand, in \ac{OFDM}-\ac{RSMA} scheme, more power is allocated to the common stream to manage the interference as \ac{ICI} increases which is compatible with \ac{MISO} \ac{BC} system under imperfect \ac{CSIT} scenario \cite{rsmaSurvey_2022}. Also, it can be seen that common stream takes less power compared to private streams in noise dominated regime because orthogonal allocation option in private streams provides more flexibility to use frequency selectivity of the channel. 

\subsubsection{Fairness Analysis}

\begin{figure}[t]
\vspace{-0.2cm}
\centering
\includegraphics[width=0.8\columnwidth]{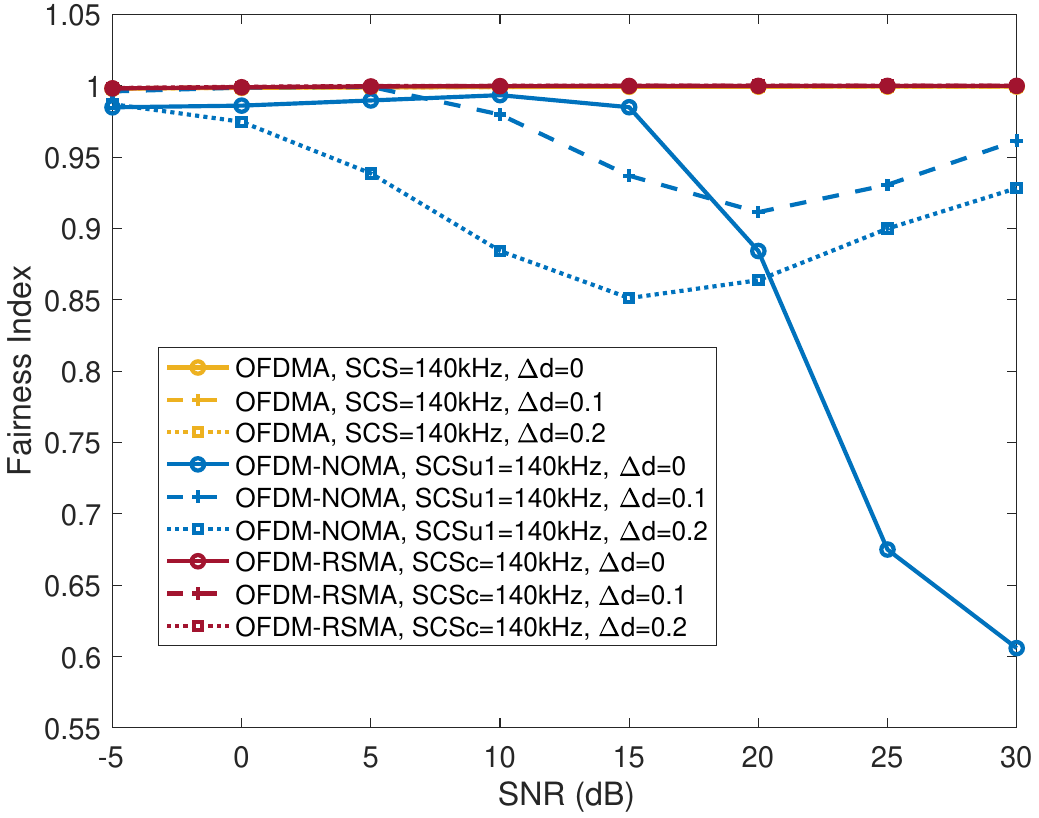}
\vspace{-0.2cm}
\caption{Fairness comparison of proposed \ac{OFDM}-\ac{RSMA} scheme with conventional \ac{OFDMA} and \ac{OFDM}-\ac{NOMA}.}
\label{fig:fairnessV2}
\vspace{-0.3cm}
\end{figure}

The fairness analysis of proposed \ac{OFDM}-\ac{RSMA} scheme comparing to \ac{OFDMA} and \ac{OFDM}-\ac{NOMA} can be seen in Fig. \ref{fig:fairnessV2}. It illustrates the Jain's fairness index versus \ac{SNR} in \SI{}{\decibel}. The fairness index, which is in the range of $[0,1]$ is defined as follows \cite{salem_2019FairnessAnalysis}: 
\vspace{-0.2cm}
\begin{IEEEeqnarray}{rCl}
    \frac{\left( \sum_{k=1}^K R_k \right)^2}{K \sum_{k=1}^K R_k^2}. \nonumber
    \vspace{-0.1cm}
\end{IEEEeqnarray}
The maximum is achieved in the fairness index when users' rates are equal. Monte Carlo simulation technique is utilized where sum-rates on different channel realization are averaged. 

It can be seen that \ac{OFDM}-\ac{NOMA} suffers to provide fairness among users for every regimes of \ac{ICI} because of block decoding property of \ac{OFDM} causing inefficient use of \ac{SIC}. In the case of frequency selective channel without any Doppler spread, the fairness index decreases in high \ac{SNR} regime because most of the power is allocated to user-2 whose subcarriers can be rescued from being in fading situation after removing the interference of user-1 via \ac{SIC}. When there is Doppler spread in the channel, more power is allocated to the user-1 whose \ac{SCS} is large to make the scheme robust against interference. On the other hand, conventional \ac{OFDMA} and proposed \ac{OFDM}-\ac{RSMA} reach the maximum fairness index which is 1 proving that \ac{OFDM}-\ac{RSMA} provides fairness among users. 

\vspace{-0.3cm}
\section{Conclusions} \label{sec:Conclusion}

In this work, we investigate the implementation and validation of \ac{RSMA} in \ac{OFDM} waveform under different channel conditions including delay and Doppler spread. Proposed \ac{OFDM}-\ac{RSMA} structure is compared with conventional \ac{OFDMA} and \ac{OFDM}-\ac{NOMA} in \ac{SISO}-\ac{BC} model. In \ac{OFDM}-\ac{RSMA}, common stream decoded by every user are transmitted on top of private streams decoded by the intended user only. Also, \ac{OFDM}-\ac{NOMA} scheme is mathematically formulated for the first time, making it suitable for the block-based decoding of \ac{OFDM} waveform. \ac{WMMSE} based algorithms are utilized to solve corresponding optimization problems for \ac{OFDM}-\ac{RSMA} and \ac{OFDM}-\ac{NOMA} techniques. Thanks to its capability for partially decoding interference and partially treating interference as noise, proposed multi-numerology \ac{OFDM}-\ac{RSMA} technique provides robustness to \ac{ICI} where conventional \ac{OFDMA} is subject to performance limitation due to the loss of orthogonality. 

For future study, we will investigate the effect of \ac{RSMA} along with the \ac{OFDM} waveform in the multiple antenna systems where combination of errors due to propagation channel such as \ac{ICI}, \ac{ISI} and \ac{CSIT} error is present. Integration of \ac{RSMA} technique with different waveforms such as \ac{OTFS}, \ac{FBMC}, and single carrier signalling will be studied to find a robust solution against interference due to combination of waveform structure and propagation channel.

\bibliographystyle{IEEEtran}
\vspace{-0.3cm}
\bibliography{reference}

\end{document}